\documentclass[aps,twocolumn,showkeys,showpacs,preprintnumbers,prd,superscriptaddress,nofootinbib,10pt]{revtex4-1}
\usepackage{graphicx,epsf,bm,amsmath,amsfonts,amssymb,epstopdf,natbib,hyperref,color,verbatim,multirow,bm}
\hypersetup{colorlinks=true,urlcolor=blue,citecolor=blue,linkcolor=blue,menucolor=blue,anchorcolor=blue,filecolor=blue}
\usepackage{appendix}
\usepackage{amsmath}
\usepackage{float}

\date{\today}

\definecolor{darkred}{rgb}{0.8, 0.0, 0.0}
\definecolor{darkgreen}{rgb}{0, 0.8, 0.0}

\def\code#1{\texttt{#1}}

\begin{document}

\title{Trinity of black hole correspondences:\\ Shadows, quasinormal modes, graybody factors, and cautionary remarks}

\author{Davide Pedrotti}
\email{davide.pedrotti-1@unitn.it}
\affiliation{Department of Physics, University of Trento, Via Sommarive 14, 38123 Povo (TN), Italy}
\affiliation{Trento Institute for Fundamental Physics and Applications (TIFPA)-INFN, Via Sommarive 14, 38123 Povo (TN), Italy}
\affiliation{School of Mathematics and Statistics, University of Sheffield, Hounsfield Road, Sheffield S3 7RH, United Kingdom \looseness=-1}

\author{Marco Calz\`{a}}
\email{marco.calza@unitn.it}
\affiliation{Department of Physics, University of Trento, Via Sommarive 14, 38123 Povo (TN), Italy}
\affiliation{Trento Institute for Fundamental Physics and Applications (TIFPA)-INFN, Via Sommarive 14, 38123 Povo (TN), Italy}

\begin{abstract}
Correspondences between apparently distant concepts are ubiquitous in theoretical physics. In the context of black holes (BHs), quasinormal modes (QNMs) were shown to be linked to both shadows, in the so-called eikonal limit, and graybody factors (GBFs), using the WKB approximation. We test the accuracy of the QNM-GBF correspondence in the context of the Hawking radiation for static and rotating black hole configurations, with particular attention to the superradiant regime. Our analysis reveals the correspondence failure to accurately reproduce the Hawking spectrum due to divergences. Furthermore, we bridge the gap between BH shadows and GBFs by drawing a correspondence between such quantities in the case of generic static and spherically symmetric spacetime configurations. The shadow-GBF correspondence is tested for some case studies, including regular BHs, and its limitations and applicability are discussed. This study opens new perspectives by introducing a new correspondence and remarking on the caution needed when considering these connections.
\end{abstract}

\maketitle

\section{Introduction}
\label{sec:introduction}

In the realm of theoretical physics, correspondences and analogies between seemingly unrelated quantities have proven to be powerful tools for deepening our understanding of physical systems. These connections, often spanning different domains, have led to breakthroughs in areas such as quantum field theory (QFT), general relativity (GR), and condensed matter physics. Notable examples include the anti-de Sitter/conformal field theory (AdS/CFT) correspondence \cite{Maldacena:1997re,Witten:1998qj,Aharony:1999ti,Ryu:2006bv}, analogue gravity \cite{Barcelo:2005fc,Barcelo:2000tg}, spontaneous symmetry breaking \cite{Baker:1962zz,Bardeen:1957kj,Nambu:1960tm,Nambu:1960xd,Nambu:1961fr,Nambu:1961tp,Weinberg:1967tq,Anderson:1963pc,Englert:1964et,Higgs:1964pj}, and renormalization group flows \cite{Pelissetto:2000ek,Berges:2000ew,Huang:2013zaa}. All of these concepts have provided valuable insight and new perspectives into the behavior of physical systems, often revealing the hidden unity between different physical phenomena. Such analogies have uncovered a wealth of unexpected connections and have provided new ways to understand familiar concepts \cite{Jona-Lasinio:2010yph}. 

The physics of black holes (BHs) offers an extremely prolific example of how analogies and correspondences have opened new and unprecedented paths in physics. In this context, the identification between quantities of different areas traces back to the analogy of the BH quantities with classical thermodynamic characteristics. Hawking first realized this connection in his seminal papers  \cite{Hawking:1974rv,Hawking:1975vcx,Gibbons:1977mu}. This groundbreaking discovery paved the way to a fertile research area, combining gravitation, QFT, and thermodynamics \cite{Witten:2024upt}, which has deeply influenced our understanding of BHs. Another celebrated example in this context is given by analog gravity. Analog gravity is based on the study of analog systems such as fluids, Bose-Einstein condensates, or optical systems to reproduce curved spacetime effects in controlled environments. The reproducibility and controllability of analog systems have allowed the study of BH horizons and Hawking radiation in laboratory settings, attracting ever increasing attention from the community. Other gravitational phenomena have been studied through analog systems, leading to important advancements both in gravitation and condensed matter physics. For a complete review, see \cite{Barcelo:2005fc}. 

Another fascinating correspondence, involving apparently unrelated BH quantities, is the more recent shadow-QNM correspondence \cite{Jusufi:2019ltj, Stefanov:2010xz,Yang:2021zqy,Pedrotti:2024znu}. At first glance, the correspondence is quite surprising, as vibrational properties of the event horizon, the quasinormal modes (QNMs) \cite{Nollert:1999ji, Kokkotas:1999bd,Berti:2009kk,Konoplya:2011qq}, are associated with its optical appearance ~\cite{Chen:2022scf,Synge:1966okc,Luminet:1979nyg,Virbhadra:1999nm,Falcke:1999pj,Narayan:2019imo,Cunha:2018acu,Dokuchaev:2019jqq,Perlick:2021aok,Wang:2022kvg,AbhishekChowdhuri:2023ekr} . However, the shadow-QNM correspondence, based on the more fundamental geometric-optics or eikonal limit \cite{Cardoso:2008bp}, has opened new and promising research avenues from both observational and theoretical perspectives. Given the numerous BH observations, coming from both gravitational \cite{KAGRA:2021vkt} and electromagnetic ~\cite{EventHorizonTelescope:2019dse,EventHorizonTelescope:2022wkp} channels, this connection has the potential to provide strong-field, multimessenger tests of fundamental physics \cite{Chen:2022nlw, Berti:2024orb, Volkel:2022khh, Volkel:2020xlc}.

Recently, a new connection among BH quantities has been proposed, namely, between QNMs and graybody factors (GBFs) \cite{Konoplya:2024lir, Konoplya:2024vuj}.  Despite QNMs and GBFs appearing as somewhat unrelated phenomena, their correspondence is actually not surprising, as both GBFs and QNMs are obtained as spectral characteristics of the same differential equation with different boundary conditions. The QNM-GBF correspondence finds potential applications in calculating the Hawking spectra of BHs and in estimating remnant parameters from gravitational wave ringdowns. Moreover, it was recently shown that GBFs enjoy a very robust stability against small deformations of the Kerr metric with respect to QNMs \cite{Oshita:2023cjz,Oshita:2024fzf}.

The purpose of this paper is twofold. First, we explicitly discuss the challenges and potential issues that arise when implementing the QNM-GBF correspondence in the context of Hawking spectra. While the correspondence offers exciting possibilities, it certainly does not come without complications. One such concern is the accuracy and reliability of the resulting GBFs, which are intrinsically less precise if compared to the ones obtained by directly solving the Teukolsky scattering problem. We carefully check whether the precision reached with the correspondence is enough to reproduce the Hawking spectrum, paying particular attention to the superradiant regime, which becomes crucial for rotating BHs.
Second, we complete the cycle of connections by building a link between BH shadows and GBFs. We draw an equation relating the opacity of a BH, the GBFs, their optical-appearance properties such as the shadow radius and Lyapunov exponent. In other words, we build, in static and spherically symmetric spacetimes, the shadow-GBF correspondence and test it on some prototypical models of regular BHs.

The paper is organized as follows. In Sec. \ref{sec:correspondences} we first present the three quantities discussed: GBFs, QNMs, and BH shadows. We then introduce the shadow-QNM and QNM-GBF correspondences, briefly discussing the steps that led to their establishment, together with their limitations and framework of applicability. A closer look at the QNM-GBF correspondence is given in Sec. \ref{sec:evaporation}. There, we careful analyse its implications in the context of the Hawking spectrum. Specifically, we assess whether the correspondence is a good tool to reproduce the Hawking spectra of a static and a rotating BH. The connection between shadows and GBFs is drawn in Sec. \ref{sec:shadowgbf}, where the correspondence is also tested for the Schwarzschild, Bardeen, and Hayward BHs. The limits of this correspondence are also discussed. Finally, in Sec. \ref{sec:conclusions}, we draw our conclusions.

We use geometrized units with $G=c=\hbar=k_B=1$, use the signature $(-,+,+,+)$, and label the QNMs overtone number, angular number, and azimuthal number as $n$, $l$, and $m$ respectively.

\section{Correspondences}
\label{sec:correspondences}
In this section, we review the state-of-the-art in BH correspondences. As a first step, we introduce the main actors: GBFs, QNMs, and shadows. Then, we review the shadow-QNM correspondence in more detail, starting from the pioneering work \cite{Ferrari:1984zz}, up to its more recent developments \cite{Jusufi:2020dhz,Yang:2021zqy,Pedrotti:2024znu}. Finally we analyze the QNM-GBF correspondence recently established in \cite{Konoplya:2024lir,Konoplya:2024vuj}. 

\subsection{GBFs, QNMs and BH shadow }

In this work we focus on a particular subset of Petrov type D metrics \cite{Kinnersley:1969zza} given by the spherically symmetric static metrics whose line element in four-dimensional Boyer-Lindquist coordinates reads \footnote{For a discussion on the Petrov classification see Refs.~\cite{Petrov:2000bs,Coley:2004jv}.} 
\begin{equation}\label{eq:metric}
	 ds^2=-f(r) dt^2+\frac{dr^2}{g(r)} +h(r) d\Omega^2\;,
\end{equation}
with $d\Omega^2=d\theta^2+\sin^2(\theta)d\varphi^2$ the metric on the 2-sphere. Furthermore, we require such line elements to be asymptotically flat by imposing
\begin{equation}\label{eq:falloffs}
	f(r)\underset{r\rightarrow+\infty}{\longrightarrow}1\;,\;\;\;\; g(r)\underset{r\rightarrow+\infty}{\longrightarrow}1,\;\quad h(r) \xrightarrow{r \to \infty} r^2\,.
\end{equation}

With the aid of the Newmann-Penrose (NP) tetrad formalism \cite{Newman:1961qr}, it is possible to write the Teukolsky equation \cite{Teukolsky:1972my,Teukolsky:1973ha} for massless test fields of different spin $s$ as a single master equation for the respective NP-scalars \cite{Arbey:2021jif}. Upon the ansatz of separability of the equation, namely
\begin{equation}
    \Upsilon_s= \sum_{l,m} e^{-i \omega t } e^{i m \phi} S^{l}_s(\theta) Z_s(r)\,,
\label{eq:upsilon}
\end{equation}
one may write its radial part in a Schr\"odinger-like form
\begin{equation} \label{eq:SchEq}
    \partial_{r_*}Z_{s} + \Bigl(\omega^2-V_{s\, l} \Bigr)Z_{s \,} =0
\end{equation}
where
\begin{equation}\label{eq:tort}
    \frac{dr_*}{dr} = \frac{1}{\sqrt{f(r) g(r)}}
\end{equation}
and
\begin{subequations}\label{eq:potentials}
\begin{align}
	&V_{0l}=\nu^0_l\frac{f}{h}+\frac{\partial_{r_*}^2\sqrt{h}}{\sqrt{h}}\,,\\
	&V_{1l}=\nu^1_l\frac{f}{h}\,,\\
	&V_{2l}=\nu^2_l\frac{f}{h}+\frac{(\partial_{r_*}h)^2}{2h^2}-\frac{\partial_{r_*}^2\sqrt{h}}{\sqrt{h}}\,,\\
	&V_{\frac12l}=\nu^{\frac12}_l\frac{f}{h} \pm \sqrt{\nu^{\frac12}_l}\,\partial_{r_*}\left( \sqrt{\frac{f}{h}} \right)\,,
\end{align}\label{pot}
\end{subequations}
with $\nu^s_l=l(l+1)-s(s-1)$ and $S^l_s(\theta)$ being the so-called spin-weighted spherical harmonics $S^s_{l,m}(\theta, \phi)=\sum S^l_s(\theta) e^{im\phi}$, reducing to the spherical harmonics $Y_{l m}$ for  $s=0$ \cite{Berti:2005gp}.

The choices (\ref{eq:falloffs}) ensures that $V_{s\,l}$ vanishes at the horizon ($r_* \rightarrow -\infty$) and at infinity ($r_* \rightarrow +\infty$); thus the asymptotic solutions read 
\begin{subequations}
\begin{align}
	&Z_s(r_*\rightarrow -\infty) = \mathfrak{a} \;e^{i \omega r_*}+\mathfrak{b} \;e^{-i \omega r_*} \label{asympt-} \,,\\
    &Z_s(r_*\rightarrow +\infty) = a \;e^{i \omega r_*}+b \;e^{-i \omega r_*} \label{asympt+}\,.
\end{align}
\end{subequations}

\subsubsection{GBFs}
In this context, to define the GBFs, we invoke purely in-going boundary conditions at the horizon for Eq.~(\ref{eq:SchEq}). Furthermore we choose a normalization of $Z_s(r)$ that is consistent with the choices made in \cite{Konoplya:2024lir,Konoplya:2024vuj}, namely we set $\mathfrak{a}=0$ and $b=1$ and rename $\mathfrak{b}=T$ and $\mathfrak{a}=R$
\begin{subequations}
\label{eq:bc_gbf}
    \begin{align}
        Z_s &= Te^{-i\omega r_*}, \qquad\quad\,\,\qquad r_* \to - \infty,\label{eq:bc_gbf1}\\
        Z_s &= e^{-i\omega r_*} + R e ^{i\omega r_*}, \qquad r_* \to + \infty.
        \label{eq:bc_gbf2}
    \end{align}
\end{subequations}
Equation~(\ref{eq:SchEq}) and the boundary conditions of Eq.~(\ref{eq:bc_gbf}) define a scattering problem and under such choices, we can identify $T$ and $R$ as the transmission and reflection coefficients, satisfying
\begin{equation}\label{eq:current_cons}
    |R|^2+|T|^2=1.
\end{equation}
In the context of Hawking evaporation \cite{Hawking:1974rv,Hawking:1975vcx}, we interpret the transmission coefficient as a measure of the deviation from a purely blackbody emission. Indeed, the \textit{graybody} factors, which are functions of $l$, $m$, $s$, and $\omega$ are defined as the square modulus of the transmission coefficient
\begin{equation}
    \Gamma_{l m}^s(\omega) \equiv |T|^2.
\end{equation}

We notice that the boundary conditions \eqref{eq:bc_gbf} describe a wave originating at spatial infinity $r_* \to \infty$ and scattering off the BH geometric potential, being partly transmitted and partly reflected. This picture, due to the symmetry of the problem and for the purpose of determining the transmission and reflection coefficients, is equivalent to the situation in which a wave is generated at the horizon and reflected /transmitted by the effective potential barrier. Speaking of Hawking evaporation the latter is a more intuitive approach to the problem, although from a mathematical point of view, the two situations are completely equivalent.

The Hawking spectrum, namely, the number of particles emitted by a BH for a given species $i$ and spin $s$, per unit time per unit frequency, is given by
\begin{eqnarray}
\frac{d^2N_i}{dtd\omega_i}=\frac{1}{2\pi}\sum_{l} (2l+1)\frac{n_i\Gamma^s_{l}(\omega)}{ e^{2\pi \omega/\kappa}\pm 1}\,,
\label{eq:HS1}
\end{eqnarray}
where $n_i$ is the number of degrees of freedom of the species $i$, $\kappa$ is the surface gravity at the event horizon, and the $\pm$ sign takes into account for the statistics of the emitted particle.
With minimal adjustments, it is possible to obtain the Hawking emission for a rotating BH,
\begin{eqnarray}
\frac{d^2N_i}{dtd\omega_i}=\frac{1}{2\pi}\sum_{l,m}\frac{n_i\Gamma^s_{l,m}(\omega)}{ e^{2\pi k/\kappa}\pm 1}\,,
\label{eq:HS2}
\end{eqnarray}
where $k = \omega - m\Omega_H$, with $\Omega_H$ the BH angular velocity of the event horizon.

\subsubsection{QNMs}
The QNM problem is formulated by imposing boundary conditions such that a wave approaching the horizon from spatial infinity is forced to go out of the causally connected domain. This means choosing $b = 0 $ and $ \mathfrak{a} = 0 $ in Eq.~(\ref{eq:SchEq}) \cite{Nollert:1999ji}.
\begin{subequations}
\label{eq:bc_qnm}
\begin{align}
        Z_s &= \mathfrak{b}\,e^{-i\omega r_*}, \qquad r_* \to - \infty,\\
        Z_s &= a\, e^{i\omega r_*} , \,\,\,\,\qquad r_* \to + \infty.
    \end{align}
\end{subequations}
Unlike the boundary conditions of the GBFs, (\ref{eq:bc_qnm}) do not guarantee the conservation of energy and therefore the described system is dissipative. Specifically, energy is lost as it propagates to infinity or falls into the BH, thereby leaving the domain of $r_*$.  Although QNMs are introduced in the context of BH perturbations, their dissipative nature distinguishes them from the normal modes of a vibrating string or membrane, which conserve energy and are described by a self-adjoint operator with a discrete spectrum and a complete set of eigenfunctions. Although a classical normal mode analysis cannot be performed in this case, QNMs can still be defined as the (incomplete) set of eigenfunctions of Eq.~(\ref{eq:SchEq}) satisfying the boundary conditions \eqref{eq:bc_qnm}, hence the term \textit{quasi}normal modes \cite{Kokkotas:1999bd,Berti:2009kk,Konoplya:2011qq}. The associated eigenvalues,  
\begin{equation}
    \omega_{\text{QNM}} = \omega_R + i\omega_I,
\end{equation}  
are the complex frequencies, where $ \omega_R $ corresponds to the oscillation frequency, and $ \omega_I $ is related to the inverse damping time. The three indices of the overtone number $ n $, the multipole number $ l $, and the azimuthal number $ m $, label these frequencies: $\omega_{\text{QNM}}= \omega_{n l m} $. \footnote{Since the majority of this work focuses on spherically symmetric black holes, the relative QNM spectra will mostly be degenerate in $ m $.} For fixed $ (l, m) $, infinitely many modes exist, ordered by the magnitude of their imaginary part and labeled by $ n $.
The fundamental mode $ n = 0 $ is the least damped. QNMs are imprinted in the so-called ringdown phase of the gravitational wave signal of binary BH coalescence and for this reason have attracted increasing attention in the last decades, thanks to the possibility of using them to test the no-hair theorem and therefore GR itself \cite{Isi:2019aib}. The oscillations of the postmerger remnant BH are typically treated in perturbation theory and are described as a superposition of QNMs. Therefore, the QNM spectrum is a fingerprint of a given BH, as, according to the no-hair theorem it should depend only on its mass and spin. This could help in challenging the theorem itself, in particular by extracting at least two complex frequencies from a gravitational wave ringdown signals and performing BH spectroscopy  \cite{Dreyer:2003bv,Berti:2005ys}.

\subsubsection{BH shadows}
The very long baseline interferometry (VLBI) images of the supermassive BHs M87* and Sagittarius A* (Sgr A*) \cite{EventHorizonTelescope:2019dse,EventHorizonTelescope:2021dqv,EventHorizonTelescope:2022wkp,EventHorizonTelescope:2022xqj}, have opened new and unprecedented avenues for testing gravity and fundamental physics in the strong field regime \cite{Vagnozzi:2019apd,Vagnozzi:2022moj,Vagnozzi:2022tba,Qi:2019zdk,Gonzalez:2023rsd,Banerjee:2022bxg,Jusufi:2020wmp,Tsupko:2019pzg,Bambi:2019tjh,Afrin:2022ztr,Khodadi:2024ubi,Vagnozzi:2020quf, Ghosh:2023kge,Ghosh:2025igz, Virbhadra:2008ws}. This discoveries have renewed interest in black hole shadows, first studied in the pioneering works of \cite{Synge:1966okc,Luminet:1979nyg, Virbhadra:1999nm,Falcke:1999pj, Virbhadra:2002ju}.

The shadow of a BH is the gravitationally lensed image of the so-called photon sphere, a region of spacetime where photons are temporarily trapped and move along unstable null geodesics around the BH \cite{Tsupko:2019pzg,Cunha:2018acu,Dokuchaev:2019jqq,Wang:2022kvg,AbhishekChowdhuri:2023ekr, Claudel:2000yi}.\footnote{As noted in \cite{Vagnozzi:2022moj}, we mention that the existence of a photon sphere is not strictly speaking necessary for a spacetime to cast a shadow, see \cite{Joshi:2020tlq}.} To help understand the physical picture, let us imagine an asymptotic observer who looks at a BH against a bright backdrop. In this setting, we can distinguish two families of photon orbits issuing from the luminous background. On the one hand, we have the so-called capture orbits, namely those falling into the BH, never reaching the observer. On the other hand, we have the scattering orbits, reaching the observer after being scattered by the BH. The set of all dark points in the observer’s sky is the BH shadow, the boundary of which is defined by light rays that are neither captured by the BH nor scattered but are temporarily trapped in the photon sphere.

We already mentioned that an infinite set of bright rings should appear in the observer’s sky, surrounding the black hole shadow. These rings, dubbed photon rings \cite{Gralla:2019xty} correspond to photons that have traveled around the black hole a certain number of times before reaching the position of the asymptotic observer. This sequence of very thin and faint rings cannot be observed with current experimental resolution and therefore appears as a single bright ring, surrounding the shadow \cite{Gan:2021xdl,daSilva:2023jxa,Olmo:2023lil}.

Following \cite{Tsupko:2019pzg}, let us define
\begin{equation}
    F(r) \equiv \sqrt{\frac{h(r)}{f(r)}},
\end{equation}
where $f$ and $h$ come from Eq.~(\ref{eq:metric}). The related photon sphere radius $r_\text{ph}$ is obtained by solving the equation
\begin{equation}
    \frac{d}{dr}\left[F^2(r_{\text{ph}})\right] = 0,
\end{equation}
which can be recast into
\begin{equation}
    h'(r_{\text{ph}})f(r_{\text{ph}}) - h(r_{\text{ph}})f'(r_{\text{ph}}) = 0,
\end{equation}
where the prime denotes a derivative with respect to $r$. Finally, the shadow radius $R_s$ is given by the gravitationally lensed photon sphere radius $r_{\text{ph}}$ and it is therefore given by \cite{Cunha:2018acu,Dokuchaev:2019jqq,Tsupko:2019pzg}
\begin{equation}\label{RS}
    R_s = \sqrt{\frac{h(r)}{f(r)}}\Bigg|_{r = r_{\text{ph}}}.
\end{equation}
This latter reduces to the well-known Schwarzschild shadow radius, $R_s = 3\sqrt{3}M$, by setting $f(r) = 1 - 2M/r $ and $h(r) = r^2$.

\subsection{Shadow-QNM correspondence}

\label{subsec:qnmshadow}
Despite being two seemingly unrelated phenomena, BH shadows and QNMs were shown to be connected in the eikonal regime \cite{Jusufi:2019ltj,Jusufi:2020dhz,Jusufi:2020mmy,Yang:2021zqy,Pedrotti:2024znu}, i.e. when the multipole number $l$ is such that $l \gg n$. 
This connection builds upon a more fundamental correspondence, linking QNMs with the unstable null geodesics orbiting the BH \cite{Ferrari:1984zz,Cardoso:2008bp}. Indeed, in the geometric optics approximation, it was shown that
\begin{equation}
    \omega_{\text{QNM}}  \overset{\underset{\mathrm{l \gg n}}{}}{=} \Omega_c l - i \left(n + \frac12\right)|\lambda|.
    \label{eq:geoemtric-optics}
\end{equation}
In \eqref{eq:geoemtric-optics}, the real and imaginary parts are given in terms of the angular velocity of the last unstable null geodesic $\Omega_c$ and the modulus of its Lyapunov exponent $|\lambda|$, which accounts for the stability of the orbit.
Equation~(\ref{eq:geoemtric-optics}) was first proved valid for the Schwarzschild BHs \cite{Ferrari:1984zz} and later generalized to generic spherically symmetric and asymptotically flat and de Sitter spacetimes \cite{Cardoso:2008bp}. 

We briefly recall the main reasoning behind Eq.~(\ref{eq:geoemtric-optics}), following the derivation of \cite{Cardoso:2008bp}. As discussed above, perturbations around spherically symmetric spacetimes can be recast in Schr\"odinger-like form as in Eq.~(\ref{eq:SchEq}). Following the steps of \cite{Schutz:1985km}, the WKB approximation leads to the so-called QNM condition 
:\begin{equation}
    \frac{\omega^2 - V_l(r_0)}{\sqrt{2\frac{d^2V_l}{dr_*}}(r_0)}= i \left(n + \frac12\right),
    \label{eq:qnm_condition}
\end{equation}
where $r_0$ is the location of the maximum of $V_l(r)$. For large values of $l$, Eq.~(\ref{pot}) reduces to
\begin{equation}
    V_l(r) \simeq f\frac{l^2}{r^2},
\end{equation}
independently from the spin value, and combined with Eq.~(\ref{eq:qnm_condition}) leads to
\begin{equation}
    \omega_{\text{QNM}} \overset{\underset{\mathrm{l \gg n}}{}}{=} l \sqrt{\frac{f_c}{r_c^2}} - i \frac{n + 1/2}{\sqrt2}\sqrt{-\frac{r_c^2}{f_c}\left(\frac{d^2}{dr_*^2}\frac{f}{r^2}\right) \Bigr\rvert_{r = rc}}.
    \label{eq:geometric-optics2}
\end{equation}
In Eq.~(\ref{eq:geometric-optics2}), the quantities labeled by $c$ are evaluated at the unstable null orbit radius. In the same general setting, it can be shown that
\begin{equation}\label{lambda}
\Omega_c = \sqrt{\frac{f_c}{r_c^2}}, \qquad
|\lambda| = \frac{1}{\sqrt2}\sqrt{-\frac{r_c^2}{f_c}\left(\frac{d^2}{dr_*^2}\frac{f}{r^2}\right) \Bigr\rvert_{r = rc}} ,
\end{equation}
therefore leading to Eq.~(\ref{eq:geoemtric-optics}). This duality offers an intriguing perspective: namely, modes featuring wavelengths much shorter than the characteristic curvature scales of the background can be visualized as trapped null particles, orbiting the BH with angular velocity $\Omega_c$ and slowly leaking out, with a timescale given by $|\lambda|$ \cite{Cardoso:2008bp}.

Despite being quite general, the correspondence was shown to break down in some special cases for gravitational perturbations and other non-minimally coupled fields, see for example the discussions in \cite{Konoplya:2017wot,Konoplya:2019hml,Glampedakis:2019dqh,Chen:2019dip,Silva:2019scu,Chen:2021cts,Li:2021mnx,Moura:2021eln,Bryant:2021xdh,Nomura:2021efi,Guo:2021enm,Konoplya:2022gjp}. Furthermore it must be noted that it is not always possible to reproduce the entire eikonal QNM spectrum using the correspondence. Specifically, the correspondence misses the QNM frequencies that by construction cannot be obtained within the WKB framework \cite{Konoplya:2022gjp}. 

The relation between BH shadows and QNMs is easily obtained by leveraging the link between unstable circular null geodesics and BH shadows. Indeed, in light of the above discussion about BH shadows, the connection of the latter with BH QNMs should come as no surprise. The correspondence is realized in static and spherically symmetric spacetimes, in the eikonal limit through the simple equation \cite{Stefanov:2010xz,Jusufi:2019ltj,Chen:2019dip,Wei:2019jve,Cvetic:2016bxi,Khanna:2016yow}:
\begin{equation}
    R_s \overset{\underset{\mathrm{l \gg n}}{}}{=} \frac{l}{\omega_{R}},
    \label{eq:qnm-shadow}
\end{equation}
where $R_s$ is the shadow radius of a spherically symmetric BH, $\omega_R$ is the real part of the QNM frequency and $l$ the corresponding multipole number. The link between shadows and QNM was finally extended to rotating BHs, even if, in this case, the connection is less straightforward and the equations are more involved. We refer to \cite{Yang:2021zqy,Jusufi:2020dhz,Pedrotti:2024znu} for more details on this topic. Before proceeding we want to stress that the shadow-QNM correspondence naturally inherits the limitations of the correspondence between QNMs and circular null geodesics, namely that it is generally valid only for test fields and it could break in some specific settings for gravitational perturbations and non-minimally coupled fields. 

Equation~(\ref{eq:qnm-shadow}) relates only the QNM real part to the BH shadow; however, its imaginary part also influences the optical appearance of the BH. Indeed, the Lyapunov exponent is linked to the QNM imaginary part in the eikonal limit via Eq.~(\ref{eq:geoemtric-optics}) and has been shown to determine the relative flux of intensity $I$ of successive rings in the photon ring structure according to \cite{Gan:2021xdl,daSilva:2023jxa,Olmo:2023lil}
\begin{equation}
    \frac{ I_{\text{n+1}}}{ I_{\text{n}}} \approx e^{-\lambda} \qquad \text{for} \;\;n \gg 1,
    \label{eq:photon_ring}
\end{equation}
where the label $n$ is referred to the $n$th photon ring. Equations \eqref{eq:qnm-shadow} and \eqref{eq:photon_ring} show how, thanks to the eikonal correspondence between QNMs and unstable circular null geodesics, the eikonal spectrum of QNMs is imprinted in the optical appearance of BHs. This opens interesting possibilities for developing new and intriguing, even if rather futuristic, multimessenger test of gravity in the strong field regime \cite{Volkel:2020xlc, Chen:2022nlw, Berti:2024orb}.

\subsection{QNM-GBF correspondence}
QNMs have recently been shown to be linked to GBFs \cite{Konoplya:2024lir,Konoplya:2024vuj}. As mentioned, QNMs and GBFs are different characteristics of the same differential equation, Eq.~(\ref{eq:SchEq}); see also \cite{Teukolsky:1972my,Teukolsky:1973ha}. The first hint of the interplay between these quantities was pointed out in \cite{Oshita:2023cjz}, where it was shown that GBFs play an important role in the estimation of the remnant parameters and, in general, in testing the no-hair theorem. In particular it was shown that for $(l,m) = (2,2)$, the BH GBF $\Gamma_{l m}$ can be imprinted on the gravitational wave spectral amplitude $|\Tilde{h}_{l m}(\omega)|$ for $\omega \gtrsim f_{l m}$, where $f_{l m}$ is the frequency of the fundamental QNM with numbers $(l,m)$, as
\begin{equation}
    |\Tilde{h}_{l m}(\omega)| \simeq c_{l m}\sqrt{1 - \Gamma_{l m}(\omega)},
\end{equation}
with $c_{l m}$ being the gravitational wave amplitude. Furthermore, other recent studies have shown that GBFs are more stable than QNMs against small metric deformations, making them interesting alternatives for BH spectroscopy. These findings show that GBFs are less prone to environmental effects and near horizon structures, possibly affecting the boundary conditions \cite{Rosato:2024arw,Oshita:2024fzf}. The link between GBFs and QNMs was achieved in the eikonal limit through the WKB method \cite{Konoplya:2024lir,Konoplya:2024vuj}. However, it was shown in the same works that the QNM-GBF correspondence holds to a great precision also for low multipole numbers $l$, see Figs. 2 and 3 therein. 

Following \cite{Konoplya:2024lir,Konoplya:2024vuj} we sketch here the derivation of the QNM-GBF correspondence. To do so we have to develop further upon the WKB method, whose working principle consists of matching the asymptotic solutions of Eq.~(\ref{eq:SchEq}), with the solution in an intermediate region, i.e., in the vicinity of the potential peak $V_l(r_0)$. The latter is typically Taylor expanded around its maximum, leading to a simplified version of Eq.~(\ref{eq:SchEq}):
\begin{equation}
     \frac{d^2}{dr_*^2}\Psi + \left[\omega^2 - V_l(r_0) - \frac12 \frac{d^2 V_l(r_0)}{dr_*^2}(r_* - r_0)^2\right] \Psi = 0.
\end{equation}
In this example, the Taylor expansion has been truncated at the second order. The general WKB formula can be written as follows \cite{Konoplya:2019hlu}: 
\begin{equation}
    \begin{split}   
        \omega^2 &= V_l(r_0) + A_2(\mathcal{K}^2) + A_4(\mathcal{K}^2) + ...\\
        & - i\mathcal{K}\sqrt{-2\frac{d^2 V_l(r_0)}{dr_*^2}}\left(1 + A_3(\mathcal{K}^2) + A_5(\mathcal{K}^2)+...\right),
        \label{eq:WKB_general1}
    \end{split}
\end{equation}
where $A_i$ are the $i$th order WKB corrections beyond the eikonal approximation and $i = 2,3,4,...$. The $A_i$ terms depend on $\mathcal{K}$, a constant whose value depends both on the boundary conditions and the derivatives of the effective potential evaluated at its maximum. For example, the QNM condition, Eq.~(\ref{eq:qnm_condition}), is recovered by keeping the first terms in \eqref{eq:WKB_general1} and $\mathcal{K} = n + \frac12$, with $n = 0,1,2,...$. In spherically symmetric backgrounds, the effective potential can be expressed as a multipole expansion:
\begin{equation}
    V_l(r) = l^2U_0(r) + l U_1(r) + U_2(r) + l^{-1}U_3(r) + ...,
\end{equation}
where $l \geq s$ and $s = 0,1/2,1,2$ is the spin of the field. With this expansion the eikonal limit of the WKB formula, namely the first real and imaginary terms of Eq.~(\ref{eq:WKB_general1}), gives 
\begin{equation}
\begin{split}\label{eq:WKB_general2}
    \omega &= \sqrt{V_l(r_0) - i \mathcal{K}\sqrt{-2\frac{d^2 V_l(r_0)}{dr_*^2}}} + \mathcal{O}(l^{-1}) \\
    &= l \sqrt{U_0(r_0)} - i \mathcal{K}
    \sqrt{\frac{-\frac{d^2 U_0(r_0)}{dr_*^2}}{2U_0(r_0)}} + \mathcal{O}(l^{-1}).
    \end{split}
\end{equation}
Given Eq.~(\ref{eq:WKB_general2}), the QNM-GBF correspondence is achieved in two steps. First, we notice that the fundamental QNMs, $\omega_{0l m}$ are recovered from Eq.~(\ref{eq:WKB_general2}) for $\mathcal{K} = \frac12$ as
\begin{equation}
    \omega_{l m n=0} = l \sqrt{U_0(r_0)} - \frac{i}{2} \sqrt{\frac{-\frac{d^2 U_0(r_0)}{dr_*^2}}{2U_0(r_0)}}.
    \label{eq:eikonal_qnm}
\end{equation}
Second, we compare the GBFs, making use of the result of  \cite{Iyer:1986np}, having the reflection and transmission coefficient of the scattering problem
\begin{equation}
    \begin{split}
        |R|^2 &= \frac{1}{1 + e^{-2\pi i \mathcal{K}}}, \qquad 0 <|R|^2 < 1,\\
        |T|^2 &= \frac{1}{1 + e^{+2\pi i \mathcal{K}}}.
        \label{eq:TR}
    \end{split}
\end{equation}
From Eq.~(\ref{eq:WKB_general2}) in the eikonal approximation, one can express $\mathcal{K}$ as a function of $\omega$ as
\begin{equation}
\begin{split}
    -i\mathcal{K} &= \frac{\omega^2 - U_l(r_0)}{\sqrt{-2\frac{d^2 U_l(r_0)}{dr_*^2}}} + \mathcal{O}\left(l^{-1}\right)\\
    & = - \frac{\omega^2 - \text{Re}\left(\omega_{n=0l m }\right)^2}{4\text{Re}\left(\omega_{n=0l m }\right)\text{Im}\left(\omega_{n=0l m }\right)} + \mathcal{O}\left(l^{-1}\right).
    \label{eq:iK}
    \end{split}
\end{equation}
where in the second equality Eq.~(\ref{eq:eikonal_qnm}) was used. Finally Eq.~(\ref{eq:iK}) can be plugged into Eq.~(\ref{eq:TR}) to get
\begin{equation}
\begin{split}
    \Gamma_l(\omega) &\equiv |T|^2 \\
    &= \left[1 + \text{exp}\left(2\pi\frac{\omega^2 - \text{Re}\left(\omega_{n=0}\right)^2}{4\text{Re}\left(\omega_{n=0}\right)\text{Im}\left(\omega_{n=0}\right)}\right)\right]^{-1} + \mathcal{O}(l^{-1}),
    \label{eq:Gamma}
\end{split}
\end{equation}
where we omitted for clarity the labels $l,m$ for the QNM frequencies. This latter equation connects GBFs and QNMs and is obtained within the WKB framework, as shown. Nonetheless, it turned out to hold also for low values of the multipole number $l$, if higher order corrections are considered. The latter are introduced at the level of Eq.~(\ref{eq:iK}) as
\begin{equation}\label{eq:gamma_corrections}
\begin{split}
    -i\mathcal{K} &= - \frac{\omega^2 - \text{Re}\left(\omega_{n=0}\right)^2}{4\text{Re}\left(\omega_{n=0}\right)\text{Im}\left(\omega_{ n=0}\right)}\\
    &+ \Delta_1 + \Delta_2 + \Delta_f + \mathcal{O}(l^{-3})
\end{split}
\end{equation}
where $\Delta_i$ for $i = \{1,2,f\}$ are functions of the fundamental QNM frequency and the first overtone $\Delta_i = \Delta_i(\omega_{n=0},\omega_{n=1})$. We refer the reader to the original papers for more details \cite{Konoplya:2024lir,Konoplya:2024vuj}.
\label{subsec:qnmgbf}

\section{QNM-GBF correspondence: implications on Hawking spectra}
\label{sec:evaporation}

In this section, we compare the Hawking spectra of Schwarzschild and Kerr BHs obtained by using the GBFs calculated both via the QNM-GBF correspondence and by numerically solving the scattering problem. We will show that the GBFs obtained
from the correspondence are unable
to properly reproduce the Hawking spectrum of a Schwarzschild BH at low energies. Furthermore, if BH rotation is considered, the GBFs from the correspondence introduce a dramatic divergence, due to their inability to account for superradiance \cite{Brito:2015oca}.

\begin{figure}[ht!]
\centering
\includegraphics[width=0.9\columnwidth]{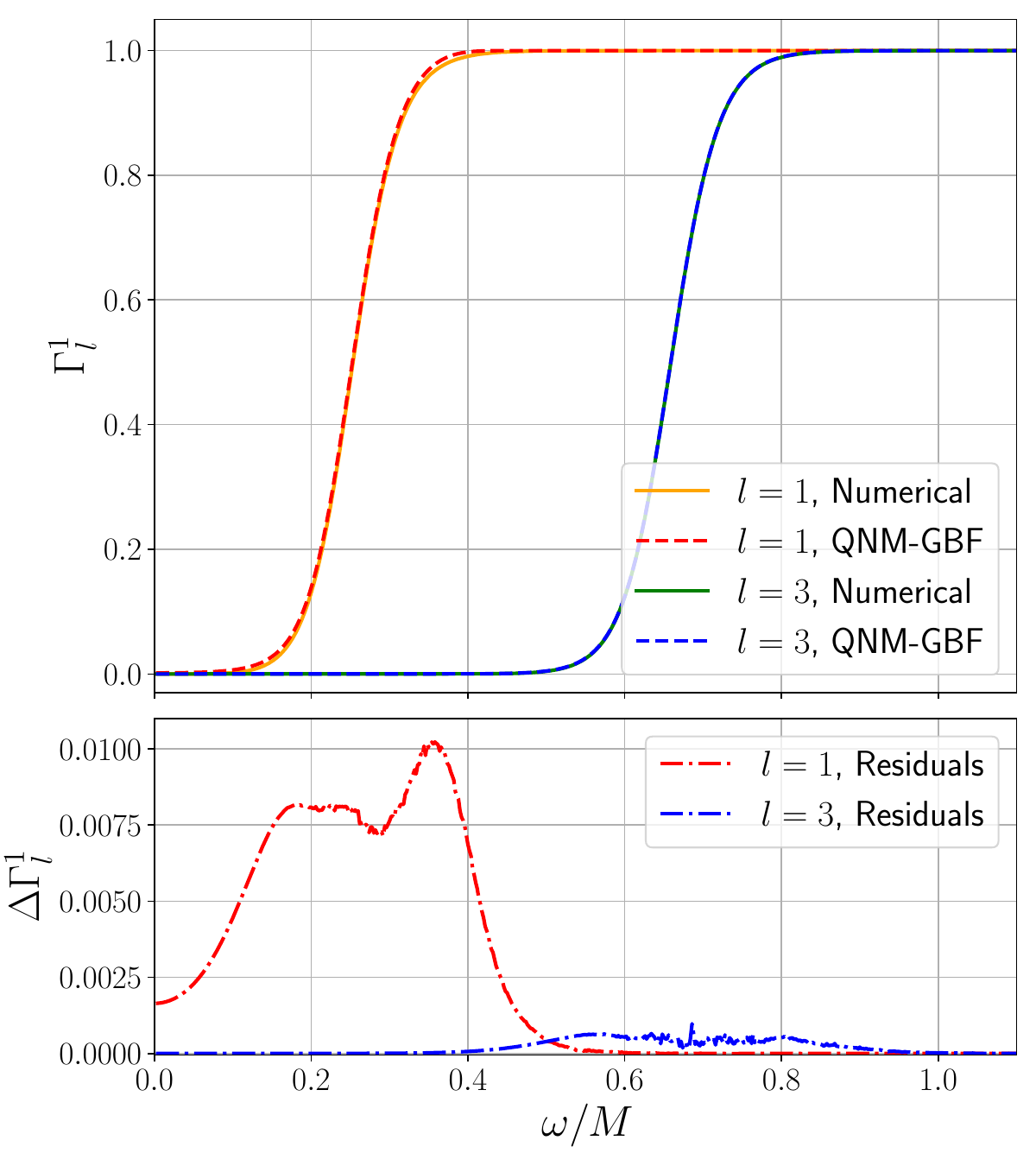}
\caption{Upper panel: comparison of the $l =1,3$ GBFs obtained by numerically solving the scattering problem (solid lines) and by using the QNM-GBF correspondence (dashed lines). In both the $l=1$ and $l=3$ cases, the GBFs computed using the two methods are visually indistinguishable. Lower panel: residuals of the GBFs computed through the correspondence against the numerically computed GBFs. In red the case $l=1$ and in blue $l=3$.}
\label{fig:GBF_comparison}
\end{figure}

Let us first focus on Schwarzschild BHs whose line element is given by Eq.~(\ref{eq:metric}), with
\begin{equation}\label{eq:Schwarzschild}
    f(r)=g(r)=\left(1- \frac{2M}{r}\right),\;\;\;\;\;\; h(r)=r^2,
\end{equation}
where $M$ is the mass of the BH.
In this case, we numerically computed the GBFs using \code{GrayHawk}, a recently released code that solves numerically the scattering problem associated with the Teukolsky equation and computes the GBFs \cite{Calza:2025whq}. Afterward, we implemented the QNM-GBF correspondence following \cite{Konoplya:2024lir}, using the sixth-order WKB formula and adopting the QNM frequencies reported on the website of Emanuele Berti ~\url{https://pages.jh.edu/eberti2/ringdown/}. In both cases, when computing the Hawking spectra, we considered a photon $(s=1)$ test field and took into account modes up to $l=4$ in Eq.~(\ref{eq:HS1}). This choice ensures adequate convergence of the summation in Eq.~(\ref{eq:potentials}), which is dominated by low $l$-modes, since the geometric potential is a growing function of $l$. It is easily noticed that the bosonic blackbody contribution of Eq.~(\ref{eq:HS1}) diverges as the frequency approaches zero. This divergence is typically offset by the GBFs in the numerator, which vanish in the same regime, thereby ensuring a finite spectrum. Even though the GBFs obtained via correspondence share this vanishing behavior, the correspondence is accurate at best at the percentage level for low $l$s, as reported in Fig.~\ref{fig:GBF_comparison}. Figure~\ref{fig:GBF_comparison} shows the comparison between the $l = 1,3$ GBFs computed with the two methods. The residuals outline how the accuracy of the correspondence increases by roughly one order of magnitude from $l = 1$ to $l=3$ modes. The wiggling behavior of the residuals is numerical noise due to the integration of the Teukolsky equation. The errors in the GBFs computed via the correspondence are substantially magnified by the divergent blackbody component, leading to inaccuracies in the low-energy region of the Hawking spectrum. Notably, low multipoles constitute the dominant contribution to the summation \eqref{eq:HS1}, making the accuracy of the corresponding GBFs particularly crucial. However, when computed using the correspondence, these low multipoles exhibit the greatest imprecision due to limitations inherent in the eikonal approximation. 

\begin{figure}[ht!]
\centering
\includegraphics[width=1.0\columnwidth]{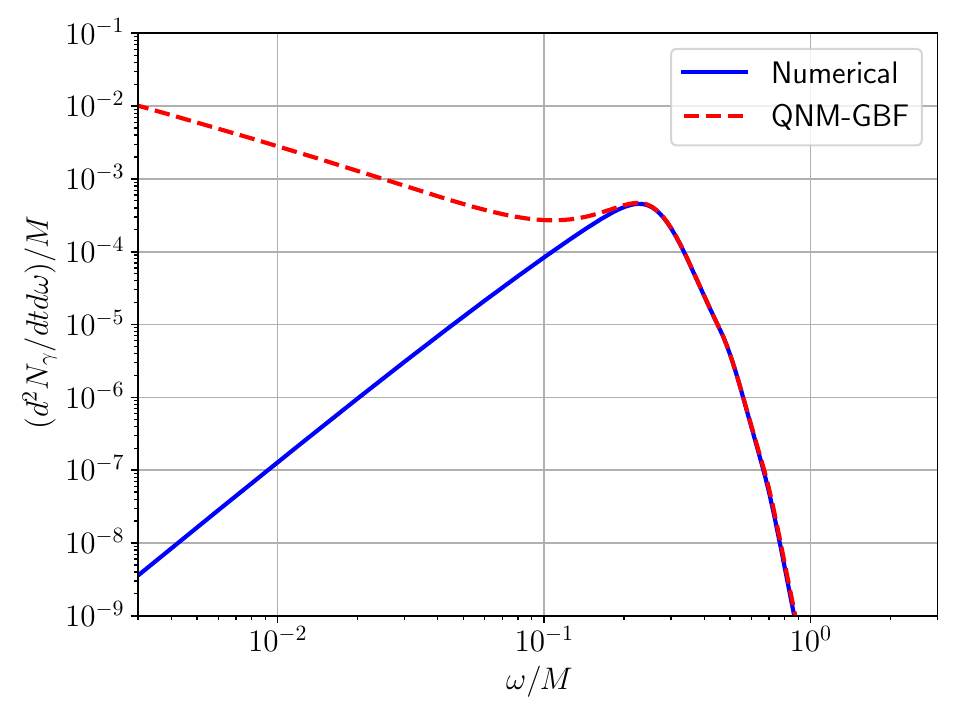}
\caption{Comparison between the mass normalized Schwarzschild 
Hawking spectrum computed with GBFs calculated solving the scattering problem (blue solid curve) and the one computed via the QNM-GBF correspondence (red dashed curve). At low frequencies, the QNM-GBF spectrum is not able to reproduce the typical infrared falloff of the Hawking spectrum.}
\label{fig:comparisonS}
\end{figure}

Figure~\ref{fig:comparisonS} shows the Hawking primary photon $(s=1)$ spectrum of a $M=1$ Schwarzschild BH, computed using both GBFs from the numerical solutions of the Teukolsky scattering problem (blue solid curve) and from the QNM-GBF correspondence (red dashed curve). It can be noticed that the correspondence fails to reproduce the numerical result in the low-energy regime, and there is a seven-orders-of-magnitude discrepancy in the lower energies considered in Fig.~\ref{fig:comparisonS}. 

Let us now focus on Kerr BHs, whose line element in Boyer-Lindquist coordinates is given by 
\begin{equation}
    \begin{split}
        ds^2 &= \frac{\Delta}{\rho^2}\left[dt - a^2\sin^2\theta d\phi\right]^2 + \\&\frac{\sin^2\theta}{\rho^2}\left[(r^2 + a^2)d\phi - a dt\right]^2 \frac{\rho^2}{\Delta}dr^2 + \rho^2 d\theta^2,
    \end{split}
\end{equation}
where $a$ is the angular momentum parameter $ 0 \leq a \leq M$ and
\begin{equation}
    \Delta = r^2 + a^2 - 2Mr, \qquad \rho^2 = r^2 + a^2\cos^2\theta.
\end{equation}
In this case, we numerically computed the GBFs following the recipe of \cite{Calza:2022ioe,Calza:2023rjt,Calza:2024fzo,Calza:2024xdh} and implemented the correspondence following \cite{Konoplya:2024vuj} under the same condition previously described for the non-rotating case.
We considered a different method to the one described in \code{GrayHawk} \cite{Calza:2025whq} since, as commented in \cite{Arbey:2025dnc}, when rotation is involved the method described in \cite{Calza:2022ioe,Calza:2023rjt,Calza:2024fzo,Calza:2024xdh} faces less numerical imprecision. We remark that the two methods provide equivalent results in the case of static BHs. 

Rotating BHs exhibit a regime in which specific bosonic perturbations are amplified, extracting energy from the BH and slowing down its rotation \cite{Brito:2015oca}. This is the so-called superradiant regime and occurs when the superradiant threshold $\omega - m\Omega_H<0$ is met. This effect leads to negative values in the GBFs for energies that satisfy the superradiant condition. As discussed in \cite{Konoplya:2024lir,Konoplya:2024vuj} the WKB approximation is such that the GBFs obtained via the correspondence are always positive, see Eq.~(\ref{eq:Gamma}), thus lacking in reproducing the correct GBFs behavior in the superradiant regime \cite{Iyer:1986np}.

\begin{figure}
    \centering
    \includegraphics[width=1\linewidth]{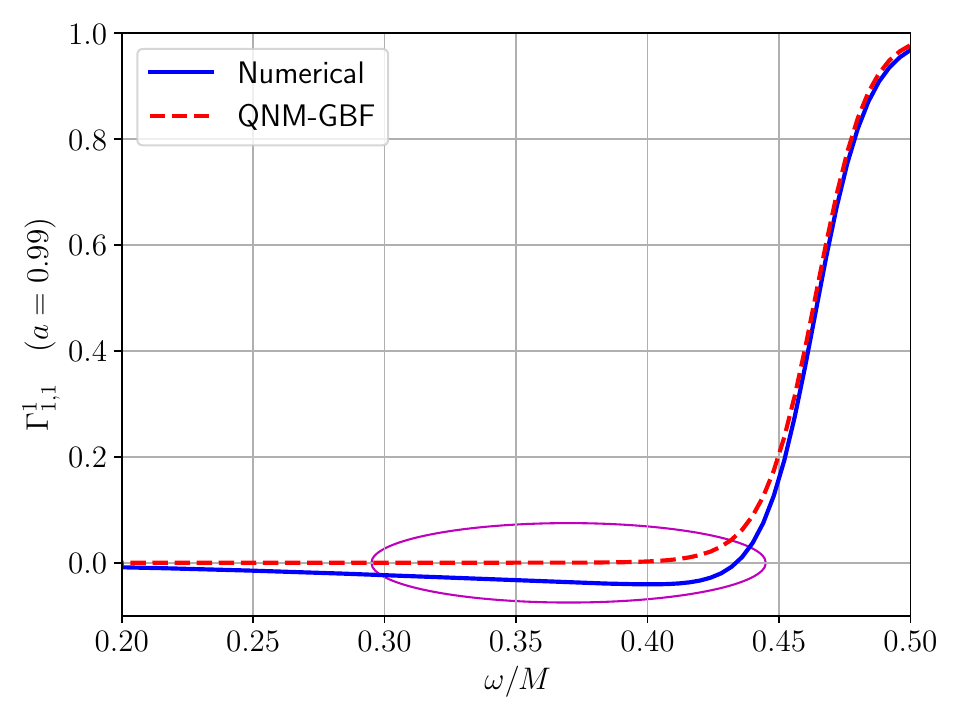}
    \caption{Comparison of the $\Gamma^{1}_{1,1}$ GBF for a nearly extremal ($a = 0.99$) Kerr BH computed by solving the Teukolsy equation (solid blue curve) and via the QNM-GBF correspondence (dashed red curve).}
    \label{fig:comparison_GBF}
\end{figure}

\begin{figure}[ht!]
    \centering
    \includegraphics[width=1\linewidth]{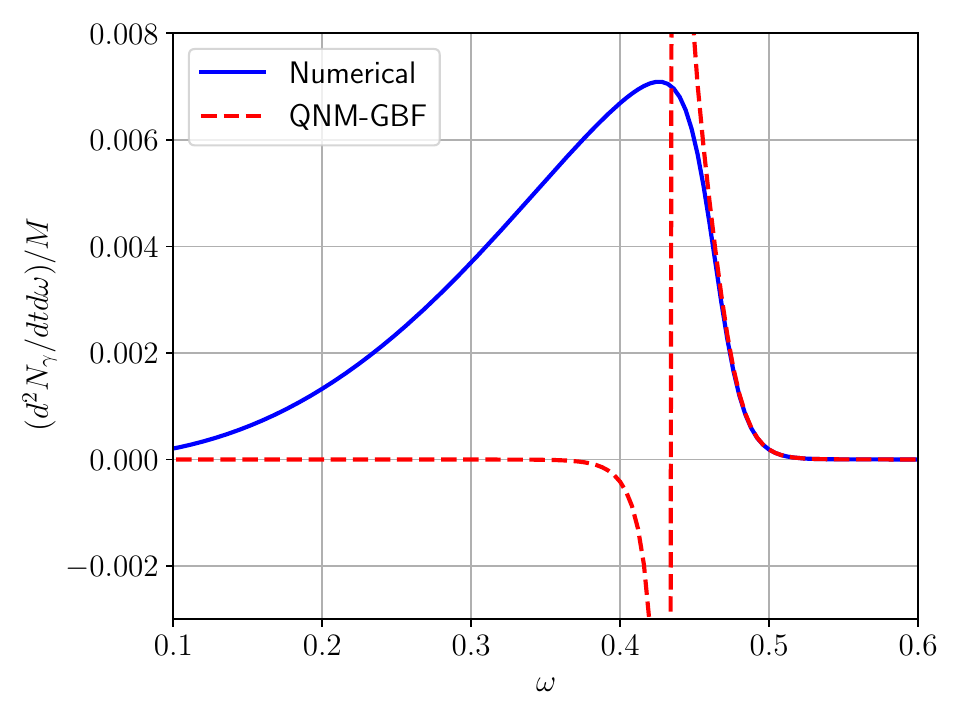}
    \caption{Comparison of the $l=1$ contribution to the Hawking spectra of a nearly extremal $(a = 0.99)$ Kerr BH, computed with GBFs obtained numerically through the scattering problem (solid blue curve) and via the QNM-GBF correspondence (red dashed curve). The QNM-GBF spectrum displays a divergence around the superradiant frequency and breaks down for frequencies $\omega < m\Omega_H$}
    \label{fig:comparisonKerrS}
\end{figure}

Figure.~\ref{fig:comparison_GBF} shows the GBF $\Gamma^{1}_{1,1}$ for a nearly extremal ($a = 0.99$) Kerr BH, computed by solving the scattering problem (solid blue) and using the correspondence (red dashed).\footnote{We considered a nearly extremal Kerr BH to emphasize the effect of superradiance, which is anyway relevant even for smaller values of $a$.} The magenta ellipse emphasizes the difference between the two lines in the superradiant regime, underlying the problematic behavior of the GBFs computed via correspondence. 
The negative sign of the GBFs undergoing superradiance is crucial in the computation of the Hawking spectrum of a rotating BH. Indeed, whenever a field mode is in the superradiant regime, the denominator in Eq.~(\ref{eq:HS2}) becomes negative and requires a negative contribution by the GBF to give an overall positive spectrum. In addition, for $\omega = m\Omega_H$ there is a divergence in the blackbody contribution to the spectrum which can be brought to a finite value only by vanishing GBFs,  behavior which cannot be provided for by GBFs obtained via the correspondence, as they are by definition positive over all the frequency domain. 
To clarify this problem, in Fig.~\ref{fig:comparisonKerrS} we compare the $l=1$ component of the Hawking spectrum of a nearly extremal $(a=0.99)$ Kerr BH of unit mass $M=1$, given by the sole $l=1$ mode computed with the two methods. As anticipated, the QNM-GBF component, plotted in Fig.~\ref{fig:comparisonKerrS}, displays a divergence around the superradiant frequency and breaks down for frequencies $\omega < m\Omega_H$.

\section{Shadow-GBF correspondence}
\label{sec:shadowgbf}

\begin{figure}[ht!]
    \centering
    \includegraphics[width=0.85\linewidth]{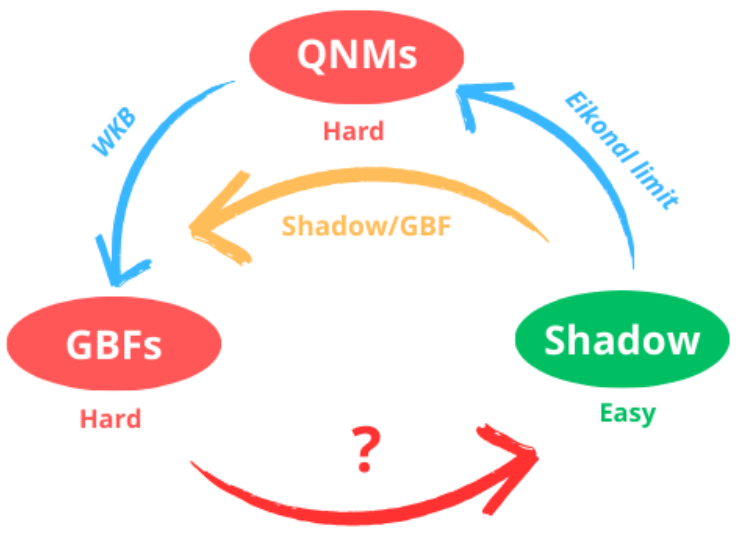}
    \caption{Graphical summary of the shadow-QNM-GBF correspondence. Blue lines represent links that have already been established and discussed in Sec. \ref{sec:correspondences}, together with the regime/framework within which they were derived. The orange arrow represents the new shadow-GBF correspondence. The color code represents the difficulty in calculating each quantity separately.}
    \label{fig:diagram}
\end{figure}

A potentially intriguing consequence of the correspondence between QNMs and GBFs is the possibility of linking the latter with the unstable circular null geodesics, and therefore the shadow, of BHs. In other words, a new, shadow-GBFs correspondence can be realized by combining the other two. The overall picture is summarized in Fig.~\ref{fig:diagram}, which connects the three quantities involved in the correspondences discussed so far: shadow, QNMs, and GBFs.
The colors qualitatively represent how difficult it is to compute each quantity: green, relatively easy to compute (BH shadows\footnote{This is true when the BH exhibits certain internal symmetries, as in the case of static, spherically symmetric and asymptotically flat spacetimes. Axisymmetric or nontrivial spacetimes have a shadow radius given analytically in terms of the BH parameters only if the spacetime allows for a separable Hamilton-Jacobi equation, or, in other words, if the spacetime possesses a Carter constant \cite{Carter:1968ks,Carter:1968rr,Chandrasekhar:1985kt}}), red, more involved to compute (QNMs and GBFs). Figure~\ref{fig:diagram} suggests the appealing possibility of computing QNMs and GBFs from the BH shadow, leveraging on the correspondences discussed earlier. The red line in Fig.~\ref{fig:diagram} labeled with a question mark, highlights our current lack of knowledge about a potential direct GBF/shadow connection. Besides the elegance of closing the triangle in Fig.~\ref{fig:diagram}, there is no solid reason to believe such a direct connection holds.

In this section, we study the shadow-GBF correspondence, introducing it and discussing its features and limitations. We consider a generic static and spherically symmetric BH in the form \eqref{eq:metric} and asymptotic behavior \eqref{eq:falloffs}. Nonetheless, we believe that the generalization to rotating BHs deserves a separate study. 

We break down the shadow-GBF connection into 3 steps:
\begin{itemize}
    \item Shadow-QNM connection,
    \item QNM-GBF connection,
    \item Synthesis: shadow-GBF connection.
\end{itemize}
The first two points have been largely discussed in Sec. \ref{sec:correspondences}; therefore we focus now on the last point.

Combining Eqs. \eqref{eq:geoemtric-optics} and \eqref{eq:Gamma} we have
    \begin{equation}
    \begin{split}
        \Gamma_{l}(\omega) &=\\
        &\left[1 + \text{exp}\left(-2\pi R_s\frac{\omega^2 - (l/R_s)^2}{2l|\lambda|}\right)\right]^{-1} + \mathcal{O}(l^{-1}).
        \end{split}
        \label{eq:shadowGBF}
    \end{equation}
Here we expressed the GBFs as a function of the shadow radius $R_s$ and the absolute value of the Lyapunov exponent $|\lambda|$.
Some comments are in order here. First, we notice that Eq.~(\ref{eq:shadowGBF}) inherits from the QNM/shadow and QNM-GBF correspondences the limitations given by the approximations considered therein. Namely, Eq.~(\ref{eq:shadowGBF}) is valid in the eikonal limit ($l \gg n$) and under the WKB approximation. Second, we stress that Eq.~(\ref{eq:shadowGBF}) is a first-order result in $l^{-1}$ and, to the best of our knowledge, it cannot be easily brought to higher orders. The reason for that is the fact that the higher order corrections $\Delta_i$ for $i = \{1,2,f\}$ to the QNM-GBF formula, Eq.~(\ref{eq:Gamma}) are functions of the differences \cite{Konoplya:2024lir,Konoplya:2024vuj}
\begin{equation}
\begin{split}
    &\text{Re}(\omega_{0 l m}) - \text{Re}(\omega_{1 l m} ) \overset{\underset{\mathrm{l \gg n}}{}}{\longrightarrow} R_s - R_s  = 0  \\
    &3\text{Im}(\omega_{0 l m}) - \text{Im}(\omega_{1 l m})  \overset{\underset{\mathrm{l \gg n}}{}}{\longrightarrow} 3 \left(|\lambda| - |\lambda|\right) = 0,
\end{split}
\end{equation}
which vanish in the eikonal limit due to Eqs. ~(\ref{eq:geoemtric-optics}) and  ~(\ref{eq:qnm-shadow}).    
As a check of the validity of Eq.~(\ref{eq:shadowGBF}), we test it for Schwarzschild, Bardeen \cite{Bardeen:1968ghw}, and Hayward \cite{Hayward:2005gi} BHs. Bardeen and Hayward BHs are among the most famous examples of BHs characterized by the absence of the central singularity being regular throughout the entire spacetime and being therefore dubbed regular BHs (RBHs). The literature on RBHs is vast, containing many proposals \cite{Borde:1996df,AyonBeato:1998ub,AyonBeato:1999rg,Bronnikov:2005gm,Berej:2006cc,Bronnikov:2012ch,Rinaldi:2012vy,Stuchlik:2014qja,Schee:2015nua,Johannsen:2015pca,Myrzakulov:2015kda,Fan:2016hvf,Sebastiani:2016ras,Toshmatov:2017zpr,Chinaglia:2017uqd,Frolov:2017dwy,Bertipagani:2020awe,Nashed:2021pah,Simpson:2021dyo,Franzin:2022iai,Chataignier:2022yic,Khodadi:2022dyi,Sajadi:2023ybm,Javed:2024wbc,Ditta:2024jrv,Al-Badawi:2024lvc,Corona:2024gth,Bueno:2024dgm,Calza:2024fzo,Calza:2024xdh}, and comprehensive reviews \cite{Ansoldi:2008jw,Nicolini:2008aj,Sebastiani:2022wbz,Torres:2022twv,Lan:2023cvz}.
In the following, we will compare the GBFs for the photon field ($s=1$) for three values of the multipole number $l=1,4,10$. We obtained the GBFs numerically using \code{GrayHawk} and through \eqref{eq:shadowGBF}, taking $R_s$ and $\lambda$ from \eqref{RS} and \eqref{lambda}.

\subsection{Schwarzschild BH}

\begin{figure*}[ht!]
\includegraphics[width=0.33\linewidth]{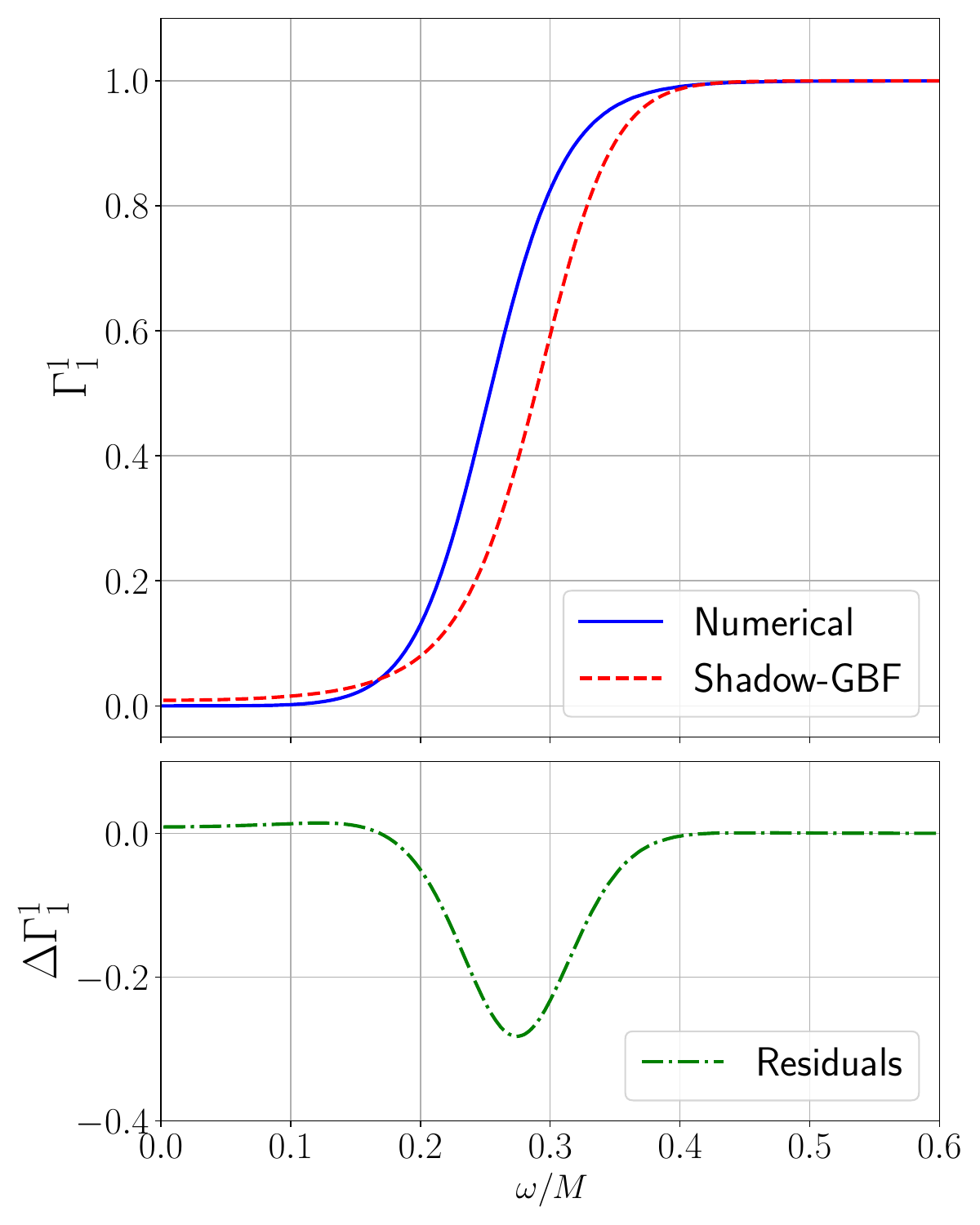}\,\,\includegraphics[width=0.33\linewidth]{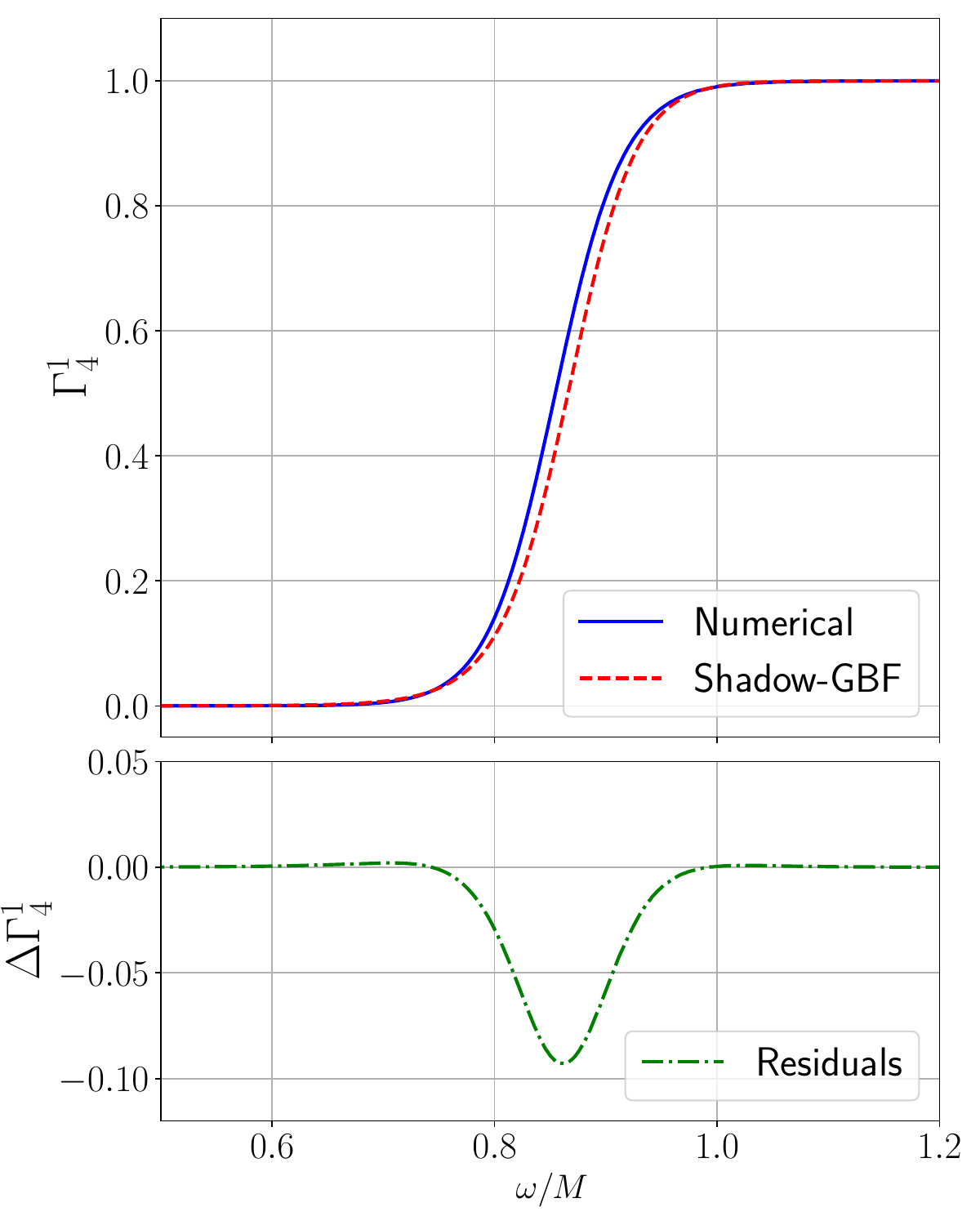}\,\,\includegraphics[width=0.33\linewidth]{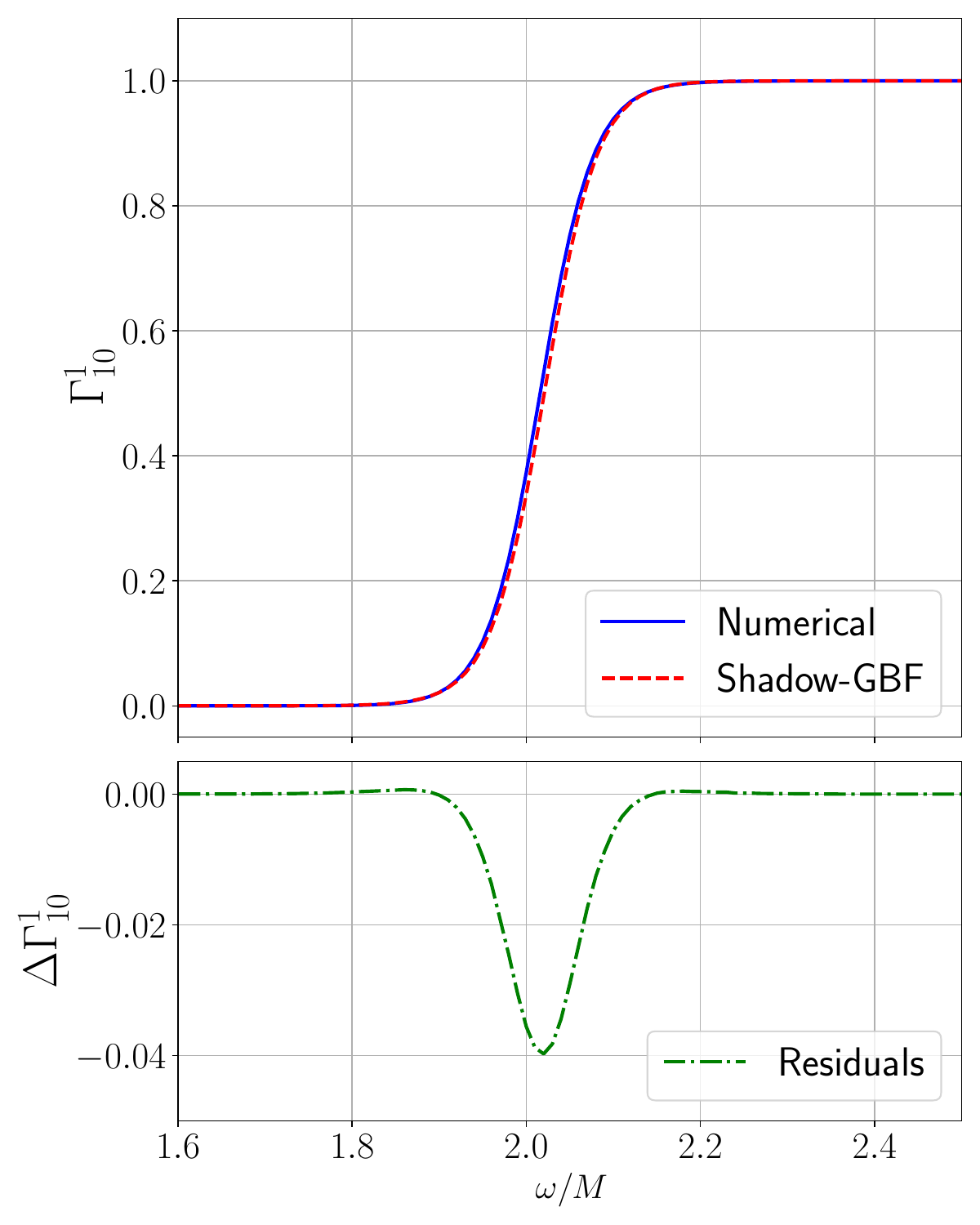}
\caption{Shadow-GBF correspondence for Schwarzschild BH for $l = 1,4,10$, with residuals. The three panels show the eikonal nature of the correspondence, which achieves a percentage-level precision only for $l = 10$.}
\label{fig:shadowGBF_S}
\end{figure*}

We start by considering the Schwarzschild BH, whose line element is given by Eq.~(\ref{eq:Schwarzschild}). In Fig.~\ref{fig:shadowGBF_S}, we propose a comparison between GBFs computed with the two methods (Teukolsky equation vs. correspondence) for $l = 1,4,10$, panels from left to right respectively. The plots show a clear trait that the correspondence is established in the eikonal limit: the match between the two curves gets better increasing the value of $l$.
Namely, we have discrepancies of roughly 30\% in the lowest mode $l=1$. This discrepancy reduces in the $l=4$ and $l=10$ GBFs, reaching roughly 10\% and 4\% respectively.
As mentioned before in this work, the lower $l$ modes contribute the most to the Hawking spectra. Therefore the GBFs obtained using this correspondence are not a good option for calculating Hawking spectra since they suffer a twofold approximation. Namely, they not only suffer from the problems discussed in Secs. \ref{sec:correspondences} C and \ref{sec:evaporation} and, deriving from the WKB approximation, but also inherit the restrictions given by the eikonal limit involved in the shadow-QNM correspondence and discussed in  Sec.\ref{sec:correspondences} B

\subsection{Bardeen BH}
\begin{figure*}[ht!]
\includegraphics[width=0.33\linewidth]{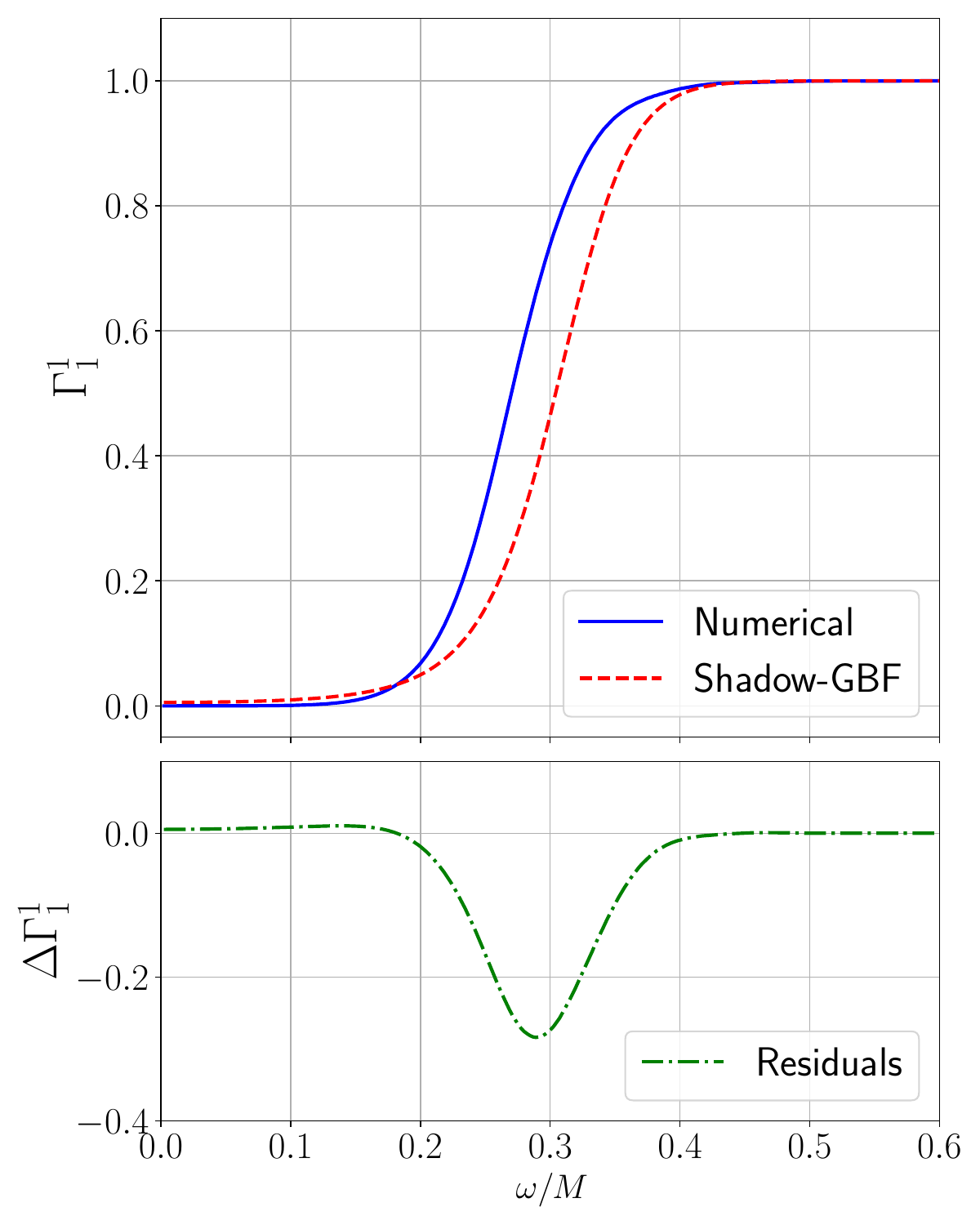}\,\,\includegraphics[width=0.33\linewidth]{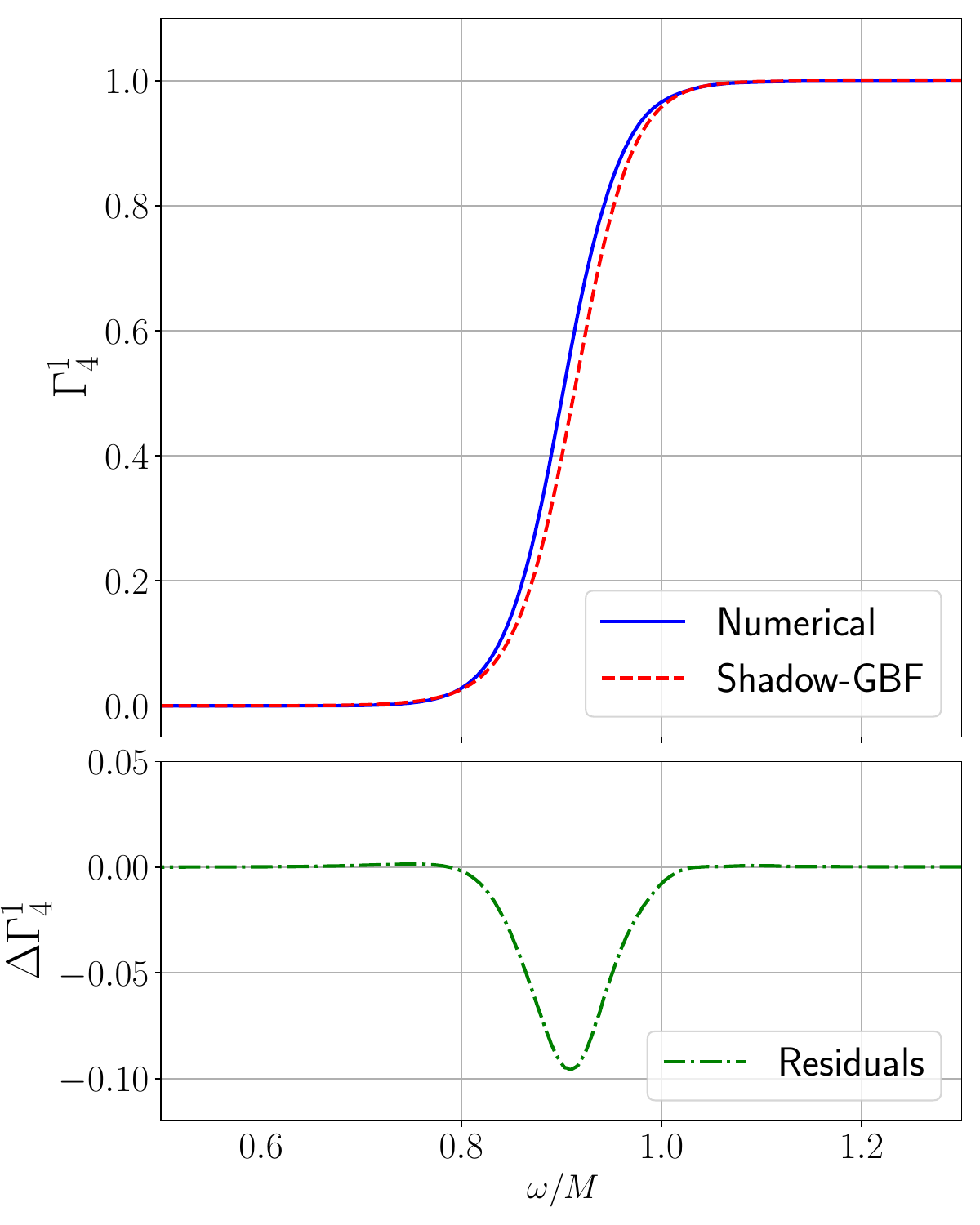}\,\,\includegraphics[width=0.33\linewidth]{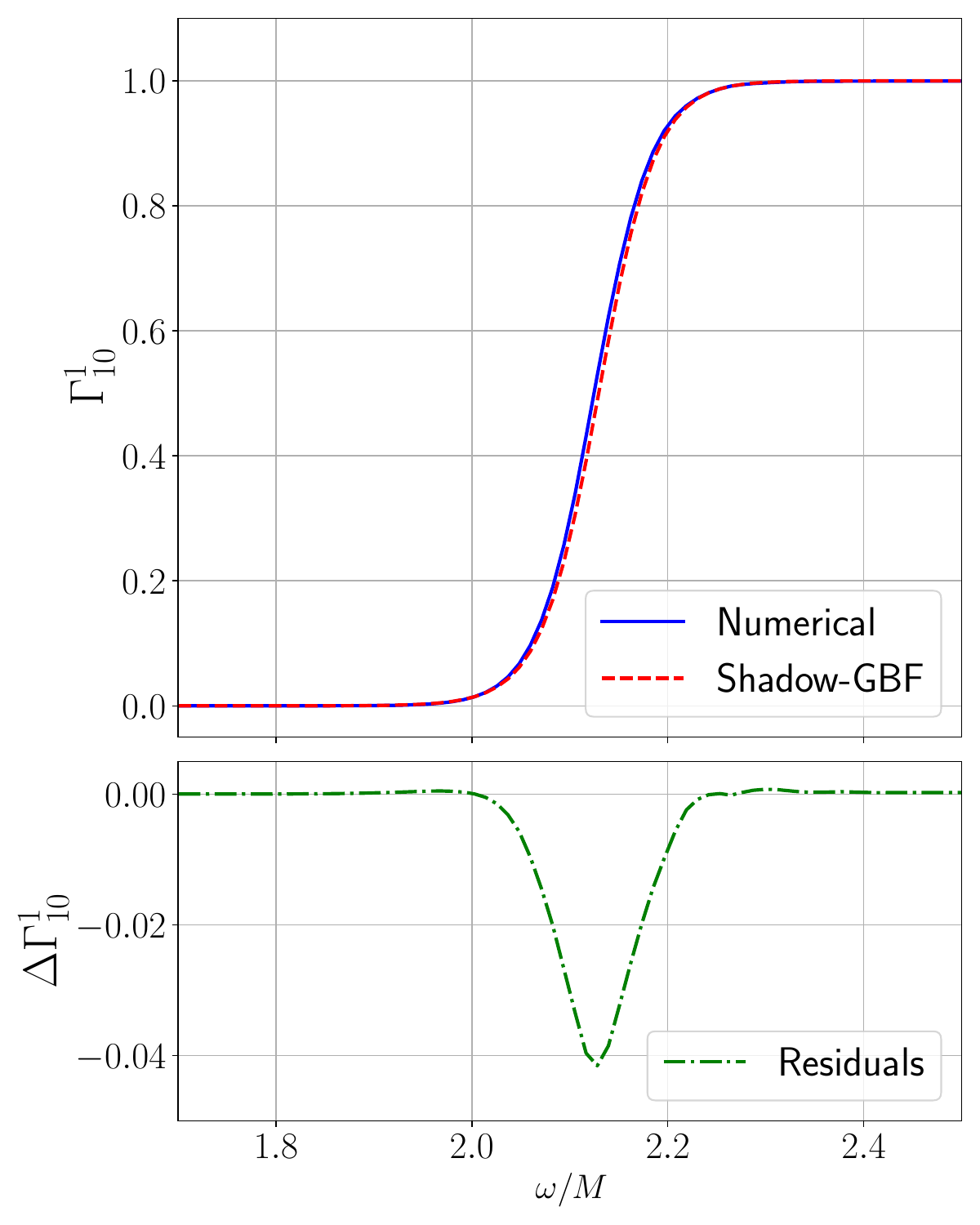}
\caption{Shadow-GBF correspondence for the Bardeen BH ($q = 0.3r_H=0.29M$) for $l = 1,4,10$, with residuals. The three panels show the eikonal nature of the correspondence, which achieves a percentage-level precision only for $l = 10$.}
\label{fig:shadowGBF_B}
\end{figure*}

The Bardeen BH~\cite{Bardeen:1968ghw} is among the first regular BHs ever proposed. Its line element in Schwarzschild coordinate is described by the following functions:
\begin{eqnarray}
f_{\text{B}}(r)=1-\frac{2Mr^2}{(r^2+q^2)^{3/2}}\quad\text{and}\quad g_{\text{B}}(r) = f_\text{B}(r)\,.
\label{eq:frbardeen}
\end{eqnarray}
Here, the parameter $q$ acts as a regularizing parameter and must satisfy the condition $q \leq \sqrt{16/27} M \sim 0.77\,M$ to ensure that spacetime describes a BH rather than a horizonless object. In the limit $q \to 0$, the Schwarzschild solution is recovered. An important feature of the Bardeen BH is the replacement of the central singularity with a de Sitter (dS) core. This behavior becomes evident as $r \to 0$, where the metric function scales as $f_B(r) \propto r^2$. Initially introduced as a phenomenological model, the Bardeen RBH has since been shown to arise from a magnetic monopole source~\cite{Ayon-Beato:2000mjt} or within the framework of non-linear electrodynamics~\cite{Ayon-Beato:2004ywd}. Additionally, quantum corrections to the uncertainty principle have been proposed as another potential origin for this RBH~\cite{Maluf:2018ksj}. Regardless of the physical mechanism behind its formation, the Bardeen solution is often treated as a phenomenological toy model, following a model-agnostic approach consistent with the treatment of other RBH spacetimes. We tested the shadow-GBF correspondence for such BH for $l = 1,4,10$ by comparing the GBFs obtained with the two methods previously discussed. In this example we took $q=0.3 r_H=0.290M$. Additionally, we checked different values of $q$, obtaining similar results. The results are plotted in Fig.~\ref{fig:shadowGBF_B}, which shows a trend very similar to the Schwarzschild case.

\subsection{Hayward BH}
A well-known example of a regular BH is the Hayward BH~\cite{Hayward:2005gi}, whose metric function is given by:  
\begin{eqnarray}  
f_{\text{H}}(r) = 1 - \frac{2Mr^2}{r^3 + 2Mq^2}\quad\text{and}\quad g_\text{H}(r) = f_\text{H}(r)\,.  
\label{eq:frhayward}  
\end{eqnarray}  
The regularizing parameter $q$ has the same upper bound of the Bardeen BH, $q \leq \sqrt{16/27}\,M$, and the Schwarzschild metric is recovered as $q \to 0$. Similarly to the Bardeen RBH, the Hayward BH replaces the central singularity with a de dS core. It was initially proposed on a phenomenological ground, despite which theoretical investigations explored potential mechanisms for its emergence, such as matter equations of state at extremely high densities~\cite{Sakharov:1966aja,1966JETP...22..378G}, finite-density and finite-curvature ~\cite{1982JETPL..36..265M,1987JETPL..46..431M,Mukhanov:1991zn}, nonlinear electrodynamics~\cite{Kumar:2020bqf,Kruglov:2021yya}, and even quantum gravity corrections~\cite{Addazi:2021xuf,AlvesBatista:2023wqm}. As with the Bardeen RBH, the Hayward solution is here treated as a model-agnostic phenomenological framework for describing singularity-free spacetimes. To further test the correspondence, we compare the GBFs of the Bardeen BH using the two methods described earlier. For this purpose, the regularization parameter is set to $q = 0.3 r_H=0.263 M$ and we considered the values $l = 1, 4, 10$. Similar results are observed for other values of $q$. Figure~\ref{fig:shadowGBF_B} shows the results highlighting a trend comparable to the Schwarzschild and Bardeen cases. Specifically, the accuracy of the correspondence improves as the eikonal limit approaches.  

\begin{figure*}[!ht]
\includegraphics[width=0.33\linewidth]{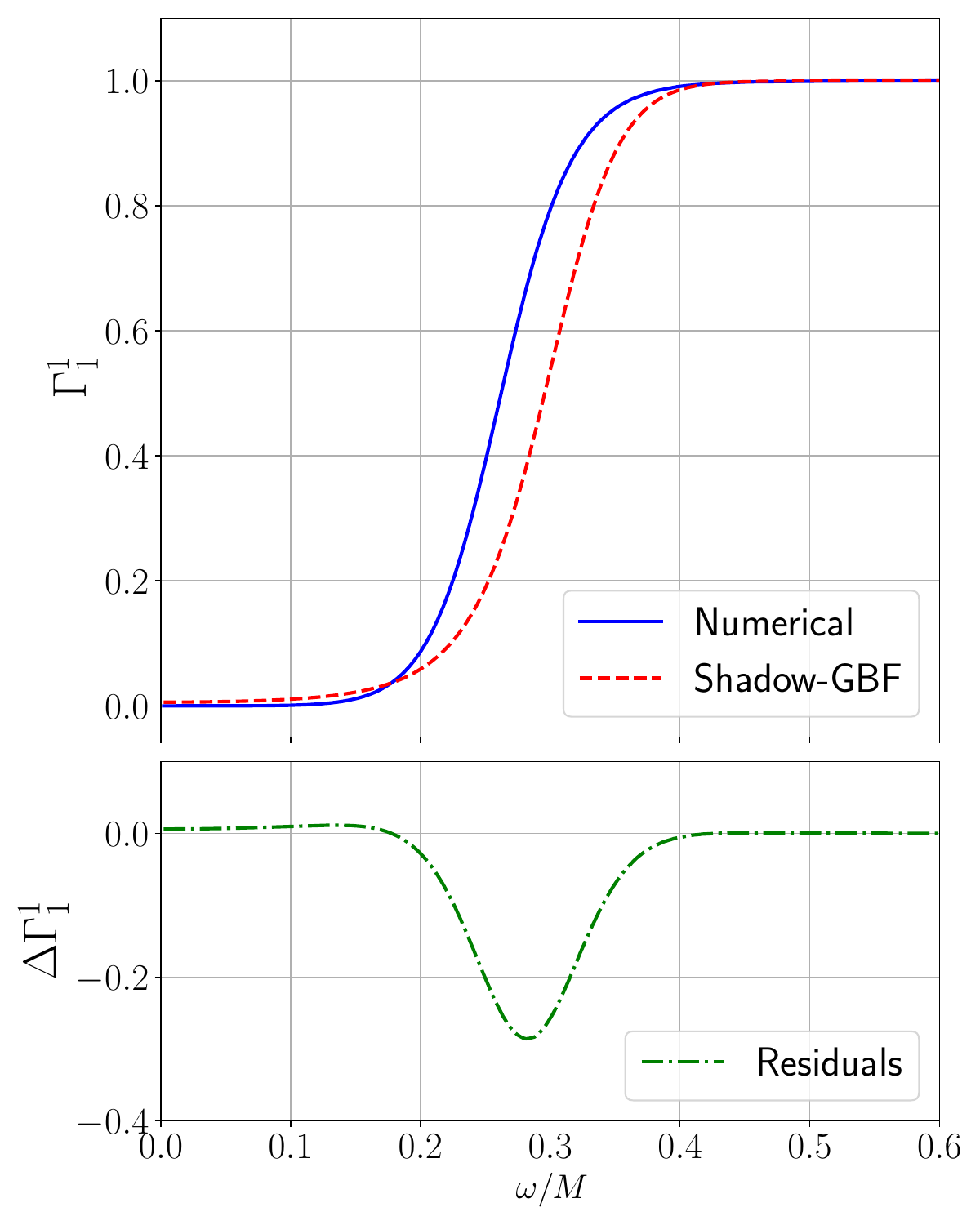}\,\,\includegraphics[width=0.33\linewidth]{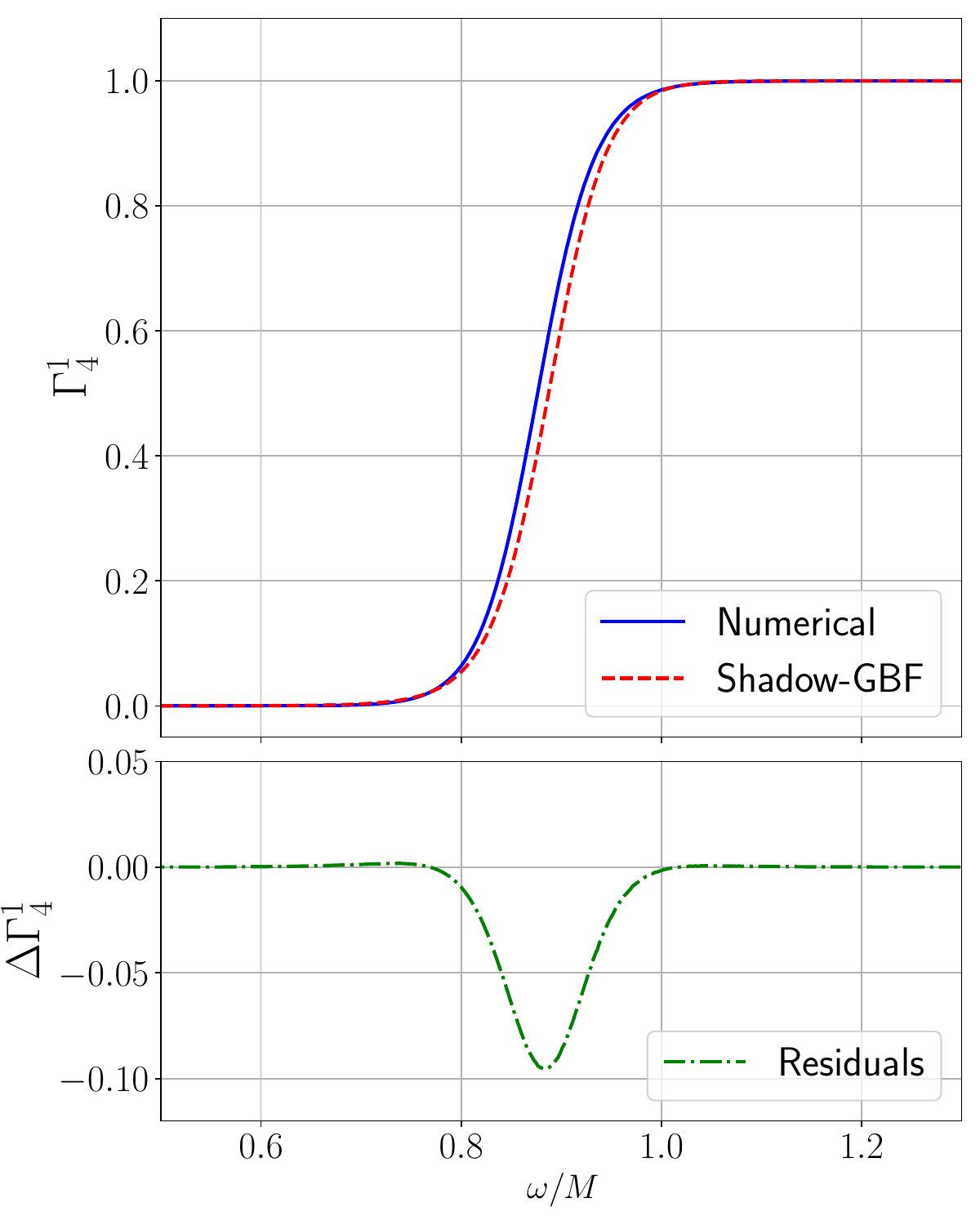}\,\,\includegraphics[width=0.33\linewidth]{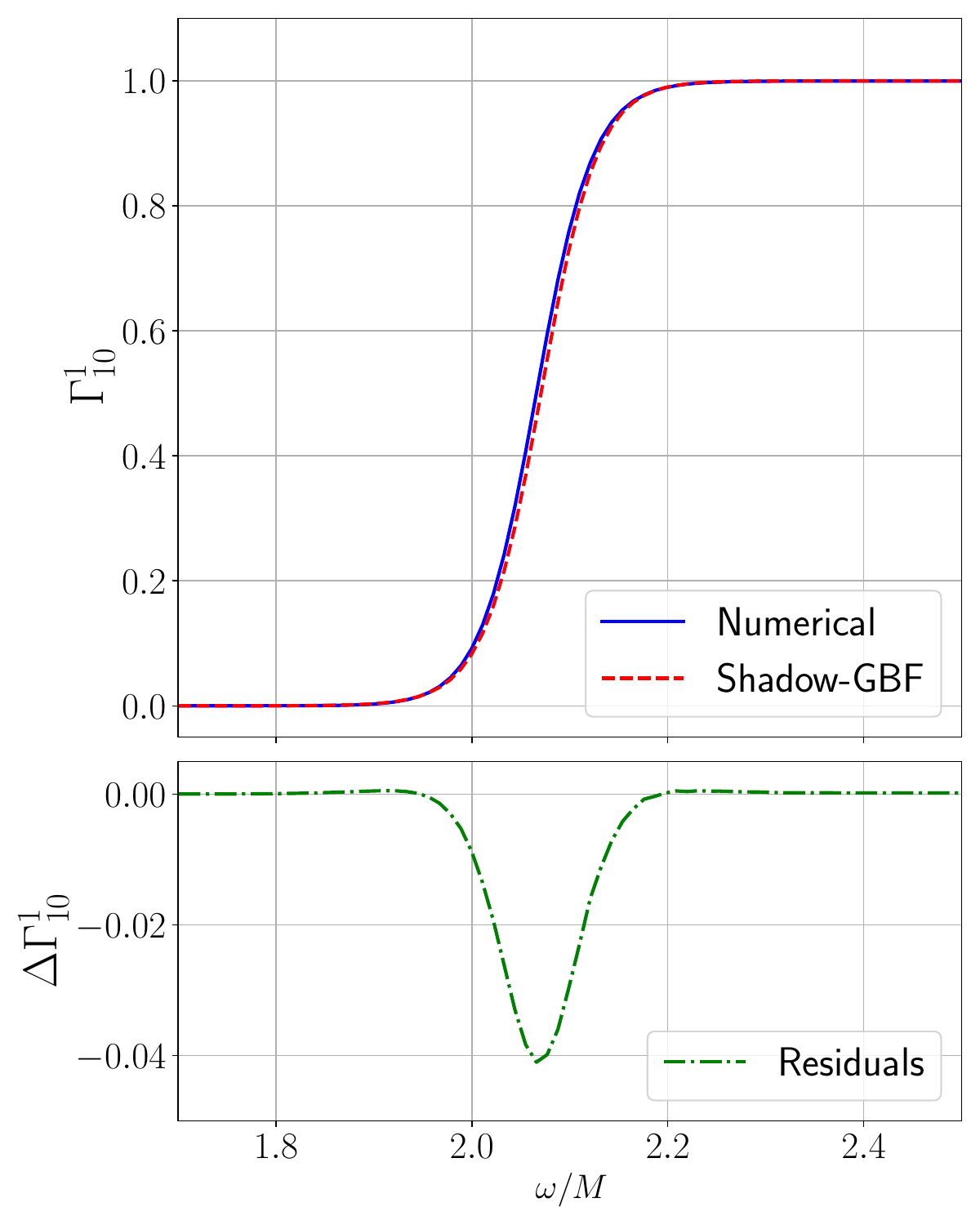}
\caption{Shadow-GBF correspondence for the Hayward BH ($q = 0.3r_H=0.263M$) for $l = 1,4,10$, with residuals. The three panels show the eikonal nature of the correspondence, which achieves a percentage-level precision only for $l = 10$.}
\label{fig:shadowGBF_H}
\end{figure*}

All three cases we have investigated have shown the success of the shadow-GBF formula \eqref{eq:shadowGBF} to reproduce the high $l$ GBFs of the BHs considered, starting from their shadow radius $R_s$ and Lyapunov exponent $\lambda$. Figures \ref{fig:shadowGBF_S}, \ref{fig:shadowGBF_B} and \ref{fig:shadowGBF_H} also clearly show the regime of applicability of the correspondence, which for all the cases considered becomes accurate at the percentage level for $l \gtrsim 10 $.


\section{Conclusions}
\label{sec:conclusions}

Over the past decade, significant progress has been made on the observational front, from the detection of gravitational wave events \cite{LIGOScientific:2018dkp,KAGRA:2021vkt} to the imaging of supermassive BH shadows at the centers of the M87 and Sgr A galaxies \cite{EventHorizonTelescope:2019dse, EventHorizonTelescope:2021dqv, EventHorizonTelescope:2022wkp,EventHorizonTelescope:2022xqj}. These observations transformed BHs from theoretical constructs into concrete physical entities, whose signatures are now routinely detected across multiple observational channels. Currently, no significant deviations from general relativity have been observed. However, the scientific community has undertaken extensive theoretical efforts to identify potential smoking guns of GR breakdown in the strong-field regime.

Motivated by such a scenario, we presented and systematically analyzed the intricate correspondences between certain characteristics of BHs, QNMs, GBFs, and shadows. We assessed the accuracy and applicability of these correspondences across different spacetimes, including Schwarzschild, Kerr, Bardeen, and Hayward metrics. We derived the shadow-GBF correspondence and focused on its validity, limitations, and implications for BH physics.

Our study reveals critical limitations that must be considered when applying these relations in astrophysical contexts. In the case of Schwarzschild BHs, we showed that while the QNM-GBF correspondence performs well at high multipoles, it fails to accurately capture the low-energy regime of the Hawking spectrum, where numerical solutions of the Teukolsky equation remain necessary. Furthermore, we reported a known result concerning Kerr BHs: the QNM-GBF correspondence breaks down in the superradiant regime, as the WKB-derived GBFs fail to account for the amplification of certain modes. We showed that this limitation incurs divergences if one tries to apply such GBFs in the computation of the Hawking spectrum, highlighting the need for refined approaches when extending this correspondence to some rotating spacetime features.

By leveraging the QNM-GBF and shadow-QNM correspondence, we built a new correspondence enabling us to infer the BH GBFs directly from its shadow properties, specifically from the shadow radius and the Lyapunov exponent. Our numerical results confirm that this relation holds with increasing accuracy as the multipole number $l$ grows, achieving a precision of a few percent for $l \gtrsim 10$. This finding not only validates the shadow-GBF connection but also reinforces the fundamental link between wave phenomena and the spacetime structure in the strong-field regime. We tested the shadow-GBF relation for  Schwarzschild, Bardeen, and Hayward BHs, and demonstrated that the correspondence retains its validity even in nonsingular spacetimes. This result suggests that the structure governing photon dynamics remains largely connected with the scattering problem in these scenarios, reinforcing the robustness of the correspondence across a wider class of BH solutions beyond general relativity’s singular Kerr and Schwarzschild ones.

From a broader perspective, our findings have significant implications for observational BH physics. The recent breakthroughs in BH imaging via the Event Horizon Telescope and the detection of GWs from binary mergers have opened new avenues for testing strong-field gravity \cite{Vagnozzi:2022moj,Berti:2018vdi, Pantig:2022qak,daSilva:2023jxa,Wei:2020ght,LIGOScientific:2018dkp}. Given that QNMs encode information about the postmerger ringdown phase of BH coalescence while shadows characterize the near-horizon region of supermassive BHs, the connection established in this paper reinforces the possibility of a unified observational framework linking gravitational wave spectroscopy with black hole imaging. This could lead to novel ways of extracting black hole parameters, such as mass, spin, charge, or more exotic ones, using a combination of multimessenger observations.

While our results provide a firm foundation for understanding the interrelations between QNMs, GBFs, and black hole shadows, several open questions remain. A natural next step would be to investigate whether these correspondences extend to higher-order corrections beyond the eikonal limit, potentially improving their accuracy for lower multipole numbers. Additionally, extending this analysis to more exotic and/or rotating spacetimes, such as higher-dimensional black holes, modified gravity solutions, and black hole mimickers, could provide further insights into the fundamental nature of these relations. An interesting avenue for future studies is creating the connection that starts from GBFs and leads to BH shadows. In this direction, studies of the inverse scattering problem \cite{10.1119/1.2190683, Lazenby:1980se, Lieb:1976cr, Volkel:2019ahb} may represent a valid help.
In conclusion, our work not only consolidates the existing correspondences between different black hole observables but also introduces and verifies a novel shadow-GBF correspondence with potential observational relevance. By bridging wave dynamics, strong-field photon trajectories, and black hole radiation properties, these findings pave the way for deeper theoretical investigations and observational applications in gravitational wave astronomy and BH imaging.

\begin{acknowledgments}
We are particularly grateful to Sunny Vagnozzi for the precious discussions.
We acknowledge support from the Istituto Nazionale di Fisica Nucleare (INFN) through the Commissione Scientifica Nazionale 4 (CSN4) Iniziativa Specifica ``Quantum Fields in Gravity, Cosmology and BHs'' (FLAG). M.C. acknowledges support from the University of Trento and the Provincia Autonoma di Trento (PAT, Autonomous Province of Trento) through the UniTrento Internal Call for Research 2023 grant ``Searching for Dark Energy off the beaten track'' (DARKTRACK, grant agreement no.\ E63C22000500003). This publication is based upon work from the COST Action CA21136 ``Addressing observational tensions in cosmology with systematics and fundamental physics'' (CosmoVerse), supported by COST (European Cooperation in Science and Technology).
\end{acknowledgments}

\bibliography{qnmgbfshadow}

\end{document}